\documentclass[twocolumn]{aastex7}
\usepackage{listings}
\providecommand{\secref}[1]{\hyperref[#1]{\S\ref{#1}}}
\providecommand{\secrefp}[1]{(\secref{#1})}

\shorttitle{The LSST Science Pipelines}
\shortauthors{Rubin Observatory Data Management Pipeline Developers}

\begin{document}

%%\doi{10.71929/rubin/2570545}
\title{The LSST Science Pipelines Software: Optical Survey Pipeline Reduction and Analysis Environment}

\AuthorCallLimit=999
% >>> input{authors} >>>
%% DO NOT EDIT THIS FILE. IT IS GENERATED FROM db2authors.py"
%% Regenerate using:
%%    python $LSST_TEXMF_DIR/bin/db2authors.py -m aas7

\author[sname='Rubin Observatory Science Pipelines Developers']{Rubin~Observatory~Science~Pipelines~Developers}
\affiliation{NSF-DOE Vera C.\ Rubin Observatory / NSF NOIRLab, 950 N.\ Cherry Ave., Tucson, AZ  85719, USA}
\email{dm-admin@lists.lsst.org}

\author[0009-0008-9216-7516,gname='Yusra',sname='AlSayyad']{Yusra~AlSayyad}
\affiliation{Department of Astrophysical Sciences, Princeton University, Princeton, NJ 08544, USA}
\email{yusra@astro.princeton.edu}

\author[0000-0002-6911-1038,gname='Robert',sname='Armstrong']{Robert~Armstrong}
\affiliation{Lawrence Livermore National Laboratory, 7000 East Avenue, Livermore, CA 94550, USA}
\email{rearmstr@gmail.com}

\author[0000-0002-5722-7199,gname='Tim S.',sname='Axelrod']{Tim~S.~Axelrod}
\affiliation{Steward Observatory, The University of Arizona, 933 N.\ Cherry Ave., Tucson, AZ 85721, USA}
\email{taxelrod@gmail.com}

\author[0000-0003-0776-8859,gname='John',sname='Banovetz']{John~Banovetz}
\affiliation{Brookhaven National Laboratory, Upton, NY 11973, USA}
\email{jdbanovetz@lbl.gov}

\author[0000-0001-8156-0429,gname='Keith',sname='Bechtol']{Keith~Bechtol}
\affiliation{NSF-DOE Vera C.\ Rubin Observatory / NSF NOIRLab, 950 N.\ Cherry Ave., Tucson, AZ  85719, USA}
\affiliation{Department of Physics, University of Wisconsin-Madison, Madison, WI 53706, USA}
\email{KBechtol@lsst.org}

\author[0000-0001-8018-5348,gname='Eric C.',sname='Bellm']{Eric~C.~Bellm}
\affiliation{University of Washington, Dept.\ of Astronomy, Box 351580, Seattle, WA 98195, USA}
\email{ecbellm@uw.edu}

\author[0000-0003-0743-9422,gname='Pedro H.',sname='Bernardinelli']{Pedro~H.~Bernardinelli}
\affiliation{Institute for Data-intensive Research in Astrophysics and Cosmology, University of Washington, 3910 15th Avenue NE, Seattle, WA 98195, USA}
\email{phbern@uw.edu}

\author[0000-0003-2759-5764,gname='James F.',sname='Bosch']{James~F.~Bosch}
\affiliation{Department of Astrophysical Sciences, Princeton University, Princeton, NJ 08544, USA}
\email{jbosch@astro.princeton.edu}

\author[0000-0002-8998-6739,gname='Andrew',sname='Bradshaw']{Andrew~Bradshaw}
\affiliation{SLAC National Accelerator Laboratory, 2575 Sand Hill Rd., Menlo Park, CA 94025, USA}
\affiliation{Kavli Institute for Particle Astrophysics and Cosmology, SLAC National Accelerator Laboratory, 2575 Sand Hill Rd., Menlo Park, CA 94025, USA}
\email{andrewkbradshaw@gmail.com}

\author[0000-0001-6966-5316,gname='Alexander',sname='Broughton']{Alexander~Broughton}
\affiliation{Kavli Institute for Particle Astrophysics and Cosmology, SLAC National Accelerator Laboratory, 2575 Sand Hill Rd., Menlo Park, CA 94025, USA}
\email{abrought@stanford.edu}

\author[0009-0003-3514-7467,gname='Audrey',sname='Budlong']{Audrey~Budlong}
\affiliation{University of Washington, Dept.\ of Physics, Box 351580, Seattle, WA 98195, USA}
\email{abudlong@uw.edu}

\author[0000-0003-3287-5250,gname='Neven',sname='Caplar']{Neven~Caplar}
\affiliation{University of Washington, Dept.\ of Astronomy, Box 351580, Seattle, WA 98195, USA}
\email{ncaplar@uw.edu}

\author[0000-0002-3936-9628,gname='Jeffrey L.',sname='Carlin']{Jeffrey~L.~Carlin}
\affiliation{NSF-DOE Vera C.\ Rubin Observatory / NSF NOIRLab, 950 N.\ Cherry Ave., Tucson, AZ  85719, USA}
\email{jeff.carlin@noirlab.edu}

\author[0009-0000-9679-3911,gname='Ross',sname='Ceballo']{Ross~Ceballo}
\affiliation{NSF-DOE Vera C.\ Rubin Observatory / NSF NOIRLab, 950 N.\ Cherry Ave., Tucson, AZ  85719, USA}
\email{ross.ceballo@noirlab.edu}

\author[0000-0001-7335-1715,gname='Colin Orion',sname='Chandler']{Colin~Orion~Chandler}
\affiliation{LSST Interdisciplinary Network for Collaboration and Computing, Tucson, USA}
\affiliation{University of Washington, Dept.\ of Astronomy, Box 351580, Seattle, WA 98195, USA}
\affiliation{Department of Astronomy and Planetary Science, Northern Arizona University, P.O.\ Box 6010, Flagstaff, AZ 86011, USA}
\email{coc123@uw.edu}

\author[0000-0002-1181-1621,gname='Hsin-Fang',sname='Chiang']{Hsin-Fang~Chiang}
\affiliation{SLAC National Accelerator Laboratory, 2575 Sand Hill Rd., Menlo Park, CA 94025, USA}
\email{hfc@stanford.edu}

\author[0000-0001-5738-8956,gname='James',sname='Chiang']{James~Chiang}
\affiliation{Kavli Institute for Particle Astrophysics and Cosmology, SLAC National Accelerator Laboratory, 2575 Sand Hill Rd., Menlo Park, CA 94025, USA}
\email{jchiang@slac.stanford.edu}

\author[0000-0003-1680-1884,gname='Yumi',sname='Choi']{Yumi~Choi}
\affiliation{NSF-DOE Vera C.\ Rubin Observatory / NSF NOIRLab, 950 N.\ Cherry Ave., Tucson, AZ  85719, USA}
\email{yumi.choi@noirlab.edu}

\author[0000-0001-9022-4232,gname='Johann',sname='Cohen-Tanugi']{Johann~Cohen-Tanugi}
\affiliation{LPCA, Universit\'{e} Clermont-Auvergne, CNRS/IN2P3, Clermont-Ferrand, France}
\email{johann.cohentanugi@gmail.com}

\author[0000-0001-5576-8189,gname='Andrew J.',sname='Connolly']{Andrew~J.~Connolly}
\affiliation{Institute for Data-intensive Research in Astrophysics and Cosmology, University of Washington, 3910 15th Avenue NE, Seattle, WA 98195, USA}
\email{ajc26@uw.edu}

\author[0000-0003-1131-7030,gname='Sylvie',sname='Dagoret-Campagne']{Sylvie~Dagoret-Campagne}
\affiliation{Universit\'{e} Paris-Saclay, CNRS/IN2P3, IJCLab, 15 Rue Georges Clemenceau, F-91405 Orsay, France}
\email{sylvie.dagoret-campagne@ijclab.in2p3.fr}

\author[0000-0003-1598-6979,gname='Gregory P.',sname='Dubois-Felsmann']{Gregory~P.~Dubois-Felsmann}
\affiliation{Caltech/IPAC, California Institute of Technology, MS 100-22, Pasadena, CA 91125-2200, USA}
\email{gpdf@ipac.caltech.edu}

\author[0000-0002-8333-7615,gname='Frossie',sname='Economou']{Frossie~Economou}
\affiliation{NSF-DOE Vera C.\ Rubin Observatory / NSF NOIRLab, 950 N.\ Cherry Ave., Tucson, AZ  85719, USA}
\email{frossie.economou@noirlab.edu}

\author[0000-0003-2933-391X,gname='Orion',sname='Eiger']{Orion~Eiger}
\affiliation{SLAC National Accelerator Laboratory, 2575 Sand Hill Rd., Menlo Park, CA 94025, USA}
\affiliation{Kavli Institute for Particle Astrophysics and Cosmology, SLAC National Accelerator Laboratory, 2575 Sand Hill Rd., Menlo Park, CA 94025, USA}
\email{eiger@slac.stanford.edu}

\author[0000-0003-3065-9941,gname='Agnès',sname='Ferté']{Agn\`{e}s~Fert\'{e}}
\affiliation{SLAC National Accelerator Laboratory, 2575 Sand Hill Rd., Menlo Park, CA 94025, USA}
\email{ferte@slac.stanford.edu}

\author[0000-0003-1898-5760,gname='Krzysztof',sname='Findeisen']{Krzysztof~Findeisen}
\affiliation{University of Washington, Dept.\ of Astronomy, Box 351580, Seattle, WA 98195, USA}
\email{kfindeis@uw.edu}

\author[0000-0001-9440-8960,gname='Merlin',sname='Fisher-Levine']{Merlin~Fisher-Levine}
\affiliation{D4D CONSULTING LTD., Suite 1 Second Floor, Everdene House, Deansleigh Road, Bournemouth, UK BH7 7DU}
\email{merlin.fisherlevin@gmail.com}

\author[0000-0001-9513-6987,gname='Michelle',sname='Gower']{Michelle~Gower}
\affiliation{NCSA, University of Illinois at Urbana-Champaign, 1205 W.\ Clark St., Urbana, IL 61801, USA}
\email{mgower@illinois.edu}

\author[0000-0003-0800-8755,gname='Leanne P.',sname='Guy']{Leanne~P.~Guy}
\affiliation{NSF-DOE Vera C.\ Rubin Observatory / NSF NOIRLab, Casilla 603, La Serena, Chile}
\email{leanne.guy@noirlab.edu}

\author[0000-0001-7203-2552,gname='Fabio',sname='Hernandez']{Fabio~Hernandez}
\affiliation{CNRS/IN2P3, CC-IN2P3, 21 avenue Pierre de Coubertin, F-69627 Villeurbanne, France}
\email{fabio@in2p3.fr}

\author[0000-0001-6718-2978,gname='Kenneth',sname='Herner']{Kenneth~Herner}
\affiliation{Fermi National Accelerator Laboratory, P. O. Box 500, Batavia, IL 60510, USA}
\email{kherner@fnal.gov}

\author[0000-0002-5292-5879,gname='Joshua',sname='Hoblitt']{Joshua~Hoblitt}
\affiliation{NSF-DOE Vera C.\ Rubin Observatory / NSF NOIRLab, 950 N.\ Cherry Ave., Tucson, AZ  85719, USA}
\email{jhoblitt@lsst.org}

\author[0000-0002-0716-947X,gname='Erin Leigh',sname='Howard']{Erin~Leigh~Howard}
\affiliation{University of Washington, Dept.\ of Astronomy, Box 351580, Seattle, WA 98195, USA}
\email{elhoward@uw.edu}

\author[0000-0003-4738-4251,gname='Minhee',sname='Hyun']{Minhee~Hyun}
\affiliation{Stanford University, 450 Jane Stanford Way, Stanford, CA 94305, USA}
\email{minhee51@stanford.edu}

\author[0009-0005-9099-4970,gname='David H.',sname='Irving']{David~H.~Irving}
\affiliation{NSF-DOE Vera C.\ Rubin Observatory / NSF NOIRLab, 950 N.\ Cherry Ave., Tucson, AZ  85719, USA}
\email{david.irving@noirlab.edu}

\author[0000-0001-5250-2633,gname='Željko',sname='Ivezić']{\v{Z}eljko~Ivezi\'{c}}
\affiliation{NSF-DOE Vera C.\ Rubin Observatory / NSF NOIRLab, 950 N.\ Cherry Ave., Tucson, AZ  85719, USA}
\affiliation{University of Washington, Dept.\ of Astronomy, Box 351580, Seattle, WA 98195, USA}
\email{ivezic@uw.edu}

\author[0000-0001-5982-167X,gname='Tim',sname='Jenness']{Tim~Jenness}
\affiliation{NSF-DOE Vera C.\ Rubin Observatory / NSF NOIRLab, 950 N.\ Cherry Ave., Tucson, AZ  85719, USA}
\email{tim.jenness@noirlab.edu}

\author[0000-0003-1996-9252,gname='Mario',sname='Jurić']{Mario~Juri\'{c}}
\affiliation{Institute for Data-intensive Research in Astrophysics and Cosmology, University of Washington, 3910 15th Avenue NE, Seattle, WA 98195, USA}
\email{mjuric@uw.edu}

\author[0000-0001-8783-6529,gname='Arun',sname='Kannawadi']{Arun~Kannawadi}
\affiliation{Department of Physics, Duke University, Durham, NC 27708, USA}
\affiliation{Department of Astrophysical Sciences, Princeton University, Princeton, NJ 08544, USA}
\email{arunkannawadi@astro.princeton.edu}

\author[0000-0001-9395-4759,gname='Lee S.',sname='Kelvin']{Lee~S.~Kelvin}
\affiliation{Department of Astrophysical Sciences, Princeton University, Princeton, NJ 08544, USA}
\email{lkelvin@astro.princeton.edu}

\author[0000-0002-4410-7868,gname='K. Simon',sname='Krughoff']{K.~Simon~Krughoff}
\altaffiliation{Author is deceased}
\affiliation{NSF-DOE Vera C.\ Rubin Observatory Project Office, 950 N.\ Cherry Ave., Tucson, AZ  85719, USA}
\email{skrughoff@lsst.org}

\author[0009-0005-5452-0671,gname='Jacob A.',sname='Kurlander']{Jacob~A.~Kurlander}
\affiliation{Institute for Data-intensive Research in Astrophysics and Cosmology, University of Washington, 3910 15th Avenue NE, Seattle, WA 98195, USA}
\email{jkurla@uw.edu}

\author[0000-0002-8357-3984,gname='Pierre-François',sname='Léget']{Pierre-Fran\c{c}ois~L\'eget}
\affiliation{Department of Astrophysical Sciences, Princeton University, Princeton, NJ 08544, USA}
\email{leget@astro.princeton.edu}

\author[0000-0002-6338-6516,gname='Kian-Tat',sname='Lim']{Kian-Tat~Lim}
\affiliation{SLAC National Accelerator Laboratory, 2575 Sand Hill Rd., Menlo Park, CA 94025, USA}
\email{ktl@slac.stanford.edu}

\author[0000-0003-1666-0962,gname='Robert H.',sname='Lupton']{Robert~H.~Lupton}
\affiliation{Department of Astrophysical Sciences, Princeton University, Princeton, NJ 08544, USA}
\email{rhl@astro.princeton.edu}

\author[0000-0002-4122-9384,gname='Nate B.',sname='Lust']{Nate~B.~Lust}
\affiliation{Department of Astrophysical Sciences, Princeton University, Princeton, NJ 08544, USA}
\email{nlust@astro.princeton.edu}

\author[0009-0003-5548-6773,gname='Lauren A.',sname='MacArthur']{Lauren~A.~MacArthur}
\affiliation{Department of Astrophysical Sciences, Princeton University, Princeton, NJ 08544, USA}
\email{lauren@astro.princeton.edu}

\author[0000-0002-9514-7245,gname='Sean Patrick',sname='MacBride']{Sean~Patrick~MacBride}
\affiliation{Physik-Institut, University of Zurich, Winterthurerstrasse 190, 8057 Zurich, Switzerland}
\email{sean.macbride@physik.uzh.ch}

\author[0009-0005-0229-7607,gname='Jeremy',sname='McCormick']{Jeremy~McCormick}
\affiliation{SLAC National Accelerator Laboratory, 2575 Sand Hill Rd., Menlo Park, CA 94025, USA}
\email{jeremym@slac.stanford.edu}

\author[0000-0001-6013-1131,gname='Guillem',sname='Megias Homar']{Guillem~Megias~Homar}
\affiliation{Division of Physics, Mathematics and Astronomy, California Institute of Technology, Pasadena, CA 91125, USA}
\email{gmegias@caltech.edu}

\author[0000-0002-2308-4230,gname='Joshua E.',sname='Meyers']{Joshua~E.~Meyers}
\affiliation{Kavli Institute for Particle Astrophysics and Cosmology, SLAC National Accelerator Laboratory, 2575 Sand Hill Rd., Menlo Park, CA 94025, USA}
\email{jmeyers314@gmail.com}

\author[0000-0001-5820-3925,gname='Joachim',sname='Moeyens']{Joachim~Moeyens}
\affiliation{Institute for Data-intensive Research in Astrophysics and Cosmology, University of Washington, 3910 15th Avenue NE, Seattle, WA 98195, USA}
\email{moeyensj@uw.edu}

\author[0000-0003-0093-4279,gname='Fred E.',sname='Moolekamp']{Fred~E.~Moolekamp}
\affiliation{soZen Inc., 105 Clearview Dr, Penfield, NY 14526}
\email{fred.moolekamp@gmail.com}

\author[0000-0002-3126-6712,gname='James R.',sname='Mullaney']{James~R.~Mullaney}
\affiliation{Astrophysics Research Cluster, School of Mathematical and Physical Sciences, University of Sheffield, Sheffield, S3 7RH, United Kingdom}
\email{j.mullaney@sheffield.ac.uk}

\author[0000-0003-3827-4691,gname='Erfan',sname='Nourbakhsh']{Erfan~Nourbakhsh}
\affiliation{Department of Astrophysical Sciences, Princeton University, Princeton, NJ 08544, USA}
\email{erfan@astro.princeton.edu}

\author[0000-0003-4141-6195,gname='William',sname='O'Mullane']{William~O'Mullane}
\affiliation{NSF-DOE Vera C.\ Rubin Observatory / NSF NOIRLab, Casilla 603, La Serena, Chile}
\email{william.omullane@noirlab.edu}

\author[0009-0001-9549-0457,gname='John K.',sname='Parejko']{John~K.~Parejko}
\affiliation{University of Washington, Dept.\ of Astronomy, Box 351580, Seattle, WA 98195, USA}
\email{parejkoj@uw.edu}

\author[0000-0002-7295-2743,gname='Hye Yun',sname='Park']{Hye~Yun~Park}
\affiliation{Department of Physics, Duke University, Durham, NC 27708, USA}
\email{hp175@duke.edu}

\author[0000-0002-4753-3387,gname='Maria T.',sname='Patterson']{Maria~T.~Patterson}
\affiliation{University of Washington, Dept.\ of Astronomy, Box 351580, Seattle, WA 98195, USA}
\email{maria.t.patterson@gmail.com}

\author[0000-0002-2158-6480,gname='Stephen R.',sname='Pietrowicz']{Stephen~R.~Pietrowicz}
\affiliation{NCSA, University of Illinois at Urbana-Champaign, 1205 W.\ Clark St., Urbana, IL 61801, USA}
\email{srp@illinois.edu}

\author[0000-0002-2598-0514,gname='Andrés A.',sname='Plazas Malagón']{Andr\'es~A.~Plazas~Malag\'on}
\affiliation{SLAC National Accelerator Laboratory, 2575 Sand Hill Rd., Menlo Park, CA 94025, USA}
\affiliation{Kavli Institute for Particle Astrophysics and Cosmology, SLAC National Accelerator Laboratory, 2575 Sand Hill Rd., Menlo Park, CA 94025, USA}
\email{plazas@stanford.edu}

\author[0000-0003-0511-0228,gname='Paul A.',sname='Price']{Paul~A.~Price}
\affiliation{Department of Astrophysical Sciences, Princeton University, Princeton, NJ 08544, USA}
\email{price@astro.princeton.edu}

\author[0000-0002-6112-9778,gname='Arianna',sname='Ranabhat']{Arianna~Ranabhat}
\affiliation{Australian Astronomical Optics, Macquarie University, North Ryde, NSW, Australia}
\email{arianna.ranabhat@mq.edu.au}

\author[0000-0003-1305-7308,gname='Meredith L.',sname='Rawls']{Meredith~L.~Rawls}
\affiliation{University of Washington, Dept.\ of Astronomy, Box 351580, Seattle, WA 98195, USA}
\affiliation{Institute for Data-intensive Research in Astrophysics and Cosmology, University of Washington, 3910 15th Avenue NE, Seattle, WA 98195, USA}
\email{mrawls@uw.edu}

\author[0000-0002-4422-0553,gname='Sophie L.',sname='Reed']{Sophie~L.~Reed}
\affiliation{Department of Astrophysical Sciences, Princeton University, Princeton, NJ 08544, USA}
\email{sophiereed@princeton.edu}

\author[0000-0001-9376-3135,gname='Eli S.',sname='Rykoff']{Eli~S.~Rykoff}
\affiliation{Kavli Institute for Particle Astrophysics and Cosmology, SLAC National Accelerator Laboratory, 2575 Sand Hill Rd., Menlo Park, CA 94025, USA}
\email{erykoff@stanford.edu}

\author[0000-0002-3623-0161,gname='Andrei',sname='Salnikov']{Andrei~Salnikov}
\affiliation{SLAC National Accelerator Laboratory, 2575 Sand Hill Rd., Menlo Park, CA 94025, USA}
\email{salnikov@slac.stanford.edu}

\author[0000-0002-8687-0669,gname='Bruno O.',sname='Sánchez']{Bruno~O.~S\'anchez}
\affiliation{Aix Marseille Universit\'{e}, CNRS/IN2P3, CPPM, 163 avenue de Luminy, F-13288 Marseille, France}
\email{bsanchez@cppm.in2p3.fr}

\author[0000-0002-4094-2102,gname='Clare',sname='Saunders']{Clare~Saunders}
\affiliation{Department of Astrophysical Sciences, Princeton University, Princeton, NJ 08544, USA}
\email{cmsaunders@princeton.edu}

\author[0000-0003-4734-2019,gname='Nima',sname='Sedaghat']{Nima~Sedaghat}
\affiliation{University of Washington, Dept.\ of Astronomy, Box 351580, Seattle, WA 98195, USA}
\email{nimaseda@uw.edu}

\author[0000-0003-3001-676X,gname='Jonathan',sname='Sick']{Jonathan~Sick}
\affiliation{J.Sick Codes Inc., Penetanguishene, Ontario, Canada}
\affiliation{NSF-DOE Vera C.\ Rubin Observatory / NSF NOIRLab, 950 N.\ Cherry Ave., Tucson, AZ  85719, USA}
\email{jsick@lsst.org}

\author[0000-0002-0558-0521,gname='Colin T.',sname='Slater']{Colin~T.~Slater}
\affiliation{University of Washington, Dept.\ of Astronomy, Box 351580, Seattle, WA 98195, USA}
\email{ctslater@uw.edu}

\author[0000-0002-3677-0571,gname='Brianna M.',sname='Smart']{Brianna~M.~Smart}
\affiliation{University of Washington, Dept.\ of Astronomy, Box 351580, Seattle, WA 98195, USA}
\email{drbsmart@uw.edu}

\author[0000-0002-2343-0949,gname='Adam',sname='Snyder']{Adam~Snyder}
\affiliation{NSF NOIRLab, 950 N.\ Cherry Ave., Tucson, AZ 85719, USA}
\email{adam.snyder@noirlab.edu}

\author[0000-0001-8708-251X,gname='Ian S.',sname='Sullivan']{Ian~S.~Sullivan}
\affiliation{University of Washington, Dept.\ of Astronomy, Box 351580, Seattle, WA 98195, USA}
\email{sullii@uw.edu}

\author[0000-0001-9445-1846,gname='John D.',sname='Swinbank']{John~D.~Swinbank}
\affiliation{ASTRON, Oude Hoogeveensedijk 4, 7991 PD, Dwingeloo, The Netherlands}
\affiliation{Department of Astrophysical Sciences, Princeton University, Princeton, NJ 08544, USA}
\email{swinbank@astron.nl}

\author[0000-0001-6268-1882,gname='Dan S.',sname='Taranu']{Dan~S.~Taranu}
\affiliation{Department of Astrophysical Sciences, Princeton University, Princeton, NJ 08544, USA}
\email{dtaranu@princeton.edu}

\author[0000-0001-9342-6032,gname='Adam J.',sname='Thornton']{Adam~J.~Thornton}
\affiliation{NSF-DOE Vera C.\ Rubin Observatory / NSF NOIRLab, 950 N.\ Cherry Ave., Tucson, AZ  85719, USA}
\email{adam.thornton@noirlab.edu}

\author[0000-0001-5538-0395,gname='Yuankun (David)',sname='Wang']{Yuankun~(David)~Wang}
\affiliation{Institute for Data-intensive Research in Astrophysics and Cosmology, University of Washington, 3910 15th Avenue NE, Seattle, WA 98195, USA}
\email{ykwang@uw.edu}

\author[0000-0003-1989-4879,gname='Christopher Z.',sname='Waters']{Christopher~Z.~Waters}
\affiliation{Department of Astrophysical Sciences, Princeton University, Princeton, NJ 08544, USA}
\email{czw@astro.princeton.edu}

\author[0000-0003-2919-7495,gname='Christina C.',sname='Williams']{Christina~C.~Williams}
\affiliation{NSF-DOE Vera C.\ Rubin Observatory / NSF NOIRLab, 950 N.\ Cherry Ave., Tucson, AZ  85719, USA}
\email{christina.williams@noirlab.edu}

\author[0000-0002-4063-883X,gname='Matthias',sname='Wittgen']{Matthias~Wittgen}
\affiliation{SLAC National Accelerator Laboratory, 2575 Sand Hill Rd., Menlo Park, CA 94025, USA}
\email{wittgen@slac.stanford.edu}

\author[0000-0001-7113-1233,gname='W. M.',sname='Wood-Vasey']{W.~M.~Wood-Vasey}
\affiliation{Department of Physics and Astronomy, University of Pittsburgh, 3941 O'Hara Street, Pittsburgh, PA 15260, USA}
\email{wmwv@pitt.edu}

\author[0000-0002-5596-198X,gname='Tianqing',sname='Zhang']{Tianqing~Zhang}
\affiliation{SLAC National Accelerator Laboratory, 2575 Sand Hill Rd., Menlo Park, CA 94025, USA}
\email{tqzhang@stanford.edu}
% <<< input{authors} <<<
\collaboration{all}{The Rubin Observatory Science Pipelines Team}
\correspondingauthor{Tim Jenness}

% >>> input{sections/abstract} >>>

\begin{abstract}

The NSF-DOE Vera C.\ Rubin Observatory is executing the Legacy Survey of Space and Time (LSST) as its prime mission,  producing a series of data releases over the ten-year survey.
The LSST Science Pipelines Software will be used to create these data releases and to perform the nightly prompt processing and alert production.
This paper provides an overview of the LSST Science Pipelines Software, describing the components and their integration into pipelines that generate science-ready data products.
We do not provide detailed descriptions of individual algorithms, but instead aim to show how the pipeline software is constructed and the breadth of the system.
Similarly, we do not provide details of specific data products and the associated file formats.

\end{abstract}
% <<< input{sections/abstract} <<<

\keywords{%
    Astrophysics - Instrumentation and Methods for Astrophysics
    ---
    methods: data analysis
    ---
    methods: miscellaneous
}

% >>> input{sections/intro} >>>
\section{Introduction}
\label{sec:intro}

The NSF-DOE Vera C.\ Rubin Observatory began performing the 10-year Legacy Survey of Space and Time \citep[LSST;][]{2019ApJ...873..111I} in 2026.
Rubin Observatory is located on Cerro Pach\'on in Chile and consists of the 8.4\,m Simonyi Survey Telescope \citep{2022SPIE12182E..0WT} with the 3.2-gigapixel LSSTCam survey camera \citep{10.71929/rubin/2571927,2024SPIE13096E..1SR} performing the main survey and the Rubin Auxiliary Telescope \citep{2020SPIE11452E..0UI} providing supplementary atmospheric calibration data using the LATISS instrument \citep{10.71929/rubin/2571930}.
The Data Management System \citep[DMS;][]{2025ASPC..538...53O} is designed to handle the flow of data from the telescope, approaching 10\,TB per night.
It produces the prompt data products, including alerts, and prepares annual data releases.
A central component of the DMS is the LSST Science Pipelines software, which provides the algorithms and frameworks required to process LSST data  and produce science-ready data products, including coadds, difference images, and catalogs.

The LSST Science Pipelines software consists of the building blocks and infrastructure required to construct high-performance pipelines to process the data from the LSST.
It is used both for Data Release Production (DRP) and Alert Production (AP).
It has been under development since at least 2004 \citep{2004AAS...20510811A} and has evolved significantly over the years as the project transitioned from prototyping \citep{2010SPIE.7740E..15A} and entered into formal construction \citep{2017ASPC..512..279J}.
The software is open-source and is designed to be usable by other optical telescopes, a capability that has been demonstrated with the processing of data from Hyper Suprime-Cam (HSC) on the Subaru Telescope in Hawaii \citep{2022PASJ...74..247A} as well as data from additional instruments used for internal testing.
There have also been plugins developed by the community for other instruments, including with non-imaging instruments.
It was used to make the first public data release from Rubin Observatory, the LSST Data Preview 1 \citep{2026AJ....171..360V,10.71929/rubin/2570308}, which consisted of data from the LSST Commissioning Camera \citep[LSSTComCam;][]{10.71929/rubin/2561361,2022SPIE12184E..0JS}, and LSST Data Preview 2 \citep[DP2;][]{RTN-115,10.71929/rubin/3382528} which consisted of data from the LSST Camera \citep{10.71929/rubin/2571927}.

In this paper we provide an overview of the software system, dividing it into four rough tiers:
\begin{itemize}
\item infrastructure consisting of low-level utility code \secrefp{sec:support}, data access, configuration, and execution frameworks \secrefp{sec:middleware}, and the core algorithmic primitives \secrefp{sec:core};
\item reusable mid-level algorithmic components \secrefp{sec:components};
\item high-level tasks and pipelines, along with some details of the algorithms specific to them \secrefp{sec:tasks-and-pipelines};
\item analysis and validation tooling \secrefp{sec:analysis}.
\end{itemize}
There are no sharp boundaries between these tiers; they are better considered to be a loose organizational aid than any kind of formal classification.
We do not include details of the science validation of the individual algorithms, and do not attempt to cover components of the system in uniform detail; instead we focus attention on those that are unique or novel or not well described elsewhere.
The other components of the LSST DMS, such as the workflow system \citep{2025ASPC..538..325G,2024EPJWC.29504026K}, the Qserv database \citep{Wang:2011:QDS:2063348.2063364,2025ASPC..538..114M} and the Rubin Science Platform \citep{LSE-319,2024ASPC..535..227O}, are not covered in this paper.
% <<< input{sections/intro} <<<
% >>> input{sections/support} >>>
\section{Fundamentals}
\label{sec:support}

The LSST Science Pipelines software is written in Python with C++ used for high-performance algorithms and for core classes that are usable in both languages.
We use Python 3 \citep[having ported from Python 2,][currently with a minimum version of Python 3.13]{2020ASPC..522..541J}, and the C++ layer can use C++20 features with \texttt{pybind11} being used to provide the interface from Python to C++.
Additionally, the C++ layer uses \texttt{ndarray} to allow seamless passing of C++ arrays to and from Python \texttt{numpy} arrays.
This compatibility with \texttt{numpy} is important in that it makes LSST data structures available to standard Python libraries such as SciPy and Astropy \citep{2016SPIE.9913E..0GJ,2018AJ....156..123A}.

Although all the software uses the \texttt{lsst} namespace, the code base is split into individual Python products in the LSST GitHub organization\footnote{All repositories can be found at \url{https://github.com/lsst} and we do not give individual URLs to each package in this paper.} that can be installed independently and which declare their own dependencies.
These dependencies are managed using the ``Extended Unix Product System'' \citep[EUPS;][]{EUPS,2018SPIE10707E..09J} where
most of the products are built using the SCons system \citep{2005Scons1377085} with LSST-specific extensions provided in the \texttt{sconsUtils} package enforcing standard build rules and creating the necessary Python package metadata files.
The naming convention is that if a package is named \texttt{some\_name} then in Python it will be imported as \texttt{lsst.some.name}.

For logging we always use standard Python logging but include an additional \texttt{VERBOSE} log level between \texttt{INFO} and \texttt{DEBUG} to provide additional non-debugging detail that can be enabled during batch processing, and also a \texttt{TRACE} log level to support very detailed debug logging.
This verbose logging is used for periodic logging where long-lived analysis tasks are required to issue a log message every 10 minutes to indicate to the batch system that they are still alive and actively performing work.
Orthogonally to the \texttt{TRACE} logging level there is also the ability to enable loggers named \texttt{TRACEn}, where \texttt{n} is an integer, and we can enable these loggers at \texttt{DEBUG} log level to provide more and more detail as the number increases.
Finally, we use \texttt{timer} loggers for each task to simplify the collection of runtime performance metrics.

For logging from C++ we use Log4CXX wrapped in the \texttt{lsst.log} package to make it look more like standard Python logging, while also supporting deferred string formatting such that log messages are only formed if the log message level is sufficient for the message to be logged.
These C++ log messages are forwarded to Python rather than being issued from an independent logging stream.
Finally, we also provide some LSST-specific exceptions that can be thrown from C++ code and caught in Python.

\begin{figure}
\plotone{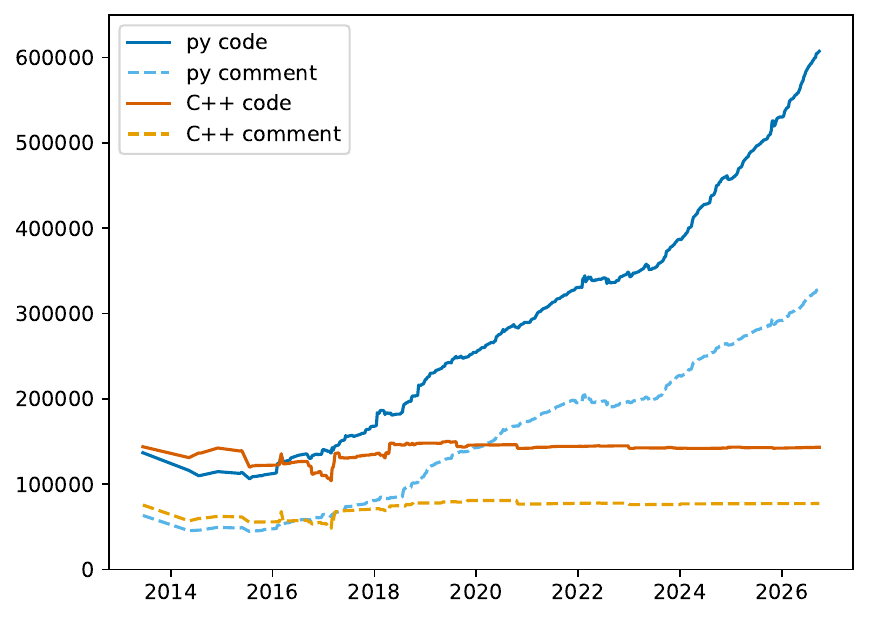}
\caption{The number of lines of code comprising the LSST Science Pipelines software as a function of year.
Line counts include comments but not blank lines. Python interfaces are implemented using \texttt{pybind11} and those are counted as C++ code. For the purposes of this count, Science Pipelines software is defined as the \texttt{lsst\_distrib} metapackage and does not include code from third-party packages or code from the Solar System Pipeline.}
\label{fig:pipe-loc}
\end{figure}

As of September 2026, the Science Pipelines software is more than 900,000 lines of Python and approximately 220,000 lines of C++.
For the purposes of this count, the Science Pipelines are defined as the \texttt{lsst\_distrib} metapackage and do not include code from third-party packages nor from the Solar System Pipeline (see \secref{sec:solsys}).
The number of lines in the pipelines code as a function of time is given in Fig.~\ref{fig:pipe-loc}.

\subsection{Python Environment}
\label{ssec:python}

An important aspect of running a large data processing campaign is to ensure that the software environment is well defined.
We define a base Python environment using conda-forge via a metapackage named \texttt{rubin-env}\footnote{\url{https://github.com/conda-forge/rubinenv-feedstock}}.
This specifies all the software needed to build and run the Science Pipelines software.
A Docker OCI image \citep{merkel2014docker} is built for each software release and the fully-specified versions of all software are recorded to ensure repeatability.

\subsection{Unit Testing and Code Coverage}

Unit testing and code coverage are critical components of code quality \citep{2018SPIE10707E..09J}.
Every package comes with unit tests written using the standard Python \texttt{unittest} module.
We run the tests using \texttt{pytest} \citep{pytest} and this comes with many advantages over our original approach of running each test file independently.
For example, running all the tests in a single process can indicate places where global state is being used unexpectedly, and randomizing the testing order can also reveal issues.
Additionally, tests can be run in parallel in multiple processes, plugins can be enabled to extend testing options and record test coverage, and a test report can be created giving details of run times and test failures.
Coding standards compliance with PEP\,8 \citep{pep8} is enforced using GitHub Actions, with \texttt{ruff} or \texttt{flake8}, and for some packages, \texttt{pre-commit} checks.

A Jenkins system \citep{2018SPIE10707E..09J,DMTN-327} provides the team with continuous integration across multiple packages.
This includes longer tests (up to a few hours) in which we run complete pipelines on small precursor datasets (typically a few GB) fetched via \texttt{git-lfs}.

\subsubsection{Installation}

Where possible, individual packages can be installed from the Python Package Index (PyPI)\footnote{For example, the Data Butler is on PyPI at \url{https://pypi.org/project/lsst-daf-butler/}. To simplify discoverability and prevent namespace confusion, Rubin packages on PyPI include the \texttt{lsst-} prefix in the package name (for example \texttt{lsst-daf-butler} and \texttt{lsst-pipe-base} corresponding to the EUPS names of \texttt{daf\_butler} and \texttt{pipe\_base}).} but that is mostly limited to packages, such as pipeline middleware packages, that do not have any C++ code due to limitations in Python packaging that prevent packages from linking directly against shared libraries from other packages.
To install the full set of pipelines software we provide an installation script that installs the appropriate Conda environment and then downloads the binaries (Linux x86\_64, Linux aarch64, or macOS Apple Silicon) from our artifact server.
As mentioned in \secref{ssec:python} it is also possible to run a Docker container with everything pre-installed.
For developers we also provide a script to build and install the software directly from the Git repositories.
% <<< input{sections/support} <<<
% >>> input{sections/middleware} >>>
\section{Data Access and Execution Abstractions}
\label{sec:middleware}

The algorithmic components of the LSST Science Pipelines are built on a suite of packages that together form a powerful data access and execution framework (\texttt{pex\_config}, \texttt{resources}, \texttt{daf\_butler}, \texttt{pipe\_base}, \texttt{ctrl\_mpexec}, and \texttt{ctrl\_bps}).
Unlike most of the rest of the codebase, these packages can be individually installed with \texttt{pip} as well as EUPS and can be used on their own.

\subsection{Data Butler}

Early in the development of the LSST Science Pipelines software it was decided that the algorithmic code should be written without knowing where files came from, what format they were written in, where the outputs are going to be written, or how they are going to be stored.
All that the algorithmic code needs to know is the relevant data model and the Python type.
To meet these requirements we developed a library called the Data Butler \citep[see e.g.,][]{2022SPIE12189E..11J,2025ASPC..538..142L}, generally referred to as simply the Butler, with the source code in a package named \texttt{daf\_butler}.

Internally, the Butler is implemented as a registry, a database keeping track of datasets, and a datastore, a storage system that can map a Butler dataset to a specific collection of bytes.
A datastore is usually a file store (including POSIX file system, S3 object stores, or WebDAV) but it is also possible to store metrics directly into the Sasquatch metrics service \citep{SQR-068,2024SPIE13101E..1MF}.

\begin{deluxetable}{ll}

%% Keep a portrait orientation

%% Over-ride the default font size
%% Use Default (12pt)

%% Use \tablewidth{?pt} to over-ride the default table width.
%% If you are unhappy with the default look at the end of the
%% *.log file to see what the default was set at before adjusting
%% this value.

%% This is the title of the table.
\tablecaption{Common dimensions present in the default dimension universe.\label{tab:dims}}

%% This command over-rides LaTeX's natural table count
%% and replaces it with this number.  LaTeX will increment
%% all other tables after this table based on this number
%% \tablenum{1}

%% The \tablehead provides the column headers.  It
%% is currently set up so that the column labels are on the
%% top line and the units surrounded by ()s are in the
%% bottom line.  You may add more header information by writing
%% another line between these lines. For each column that requires
%% extra information be sure to include a \colhead{text} command
%% and remember to end any extra lines with \\ and include the
%% correct number of &s.
\tablehead{\colhead{Name} & \colhead{Description} \\
\colhead{} & \colhead{} }

%% All data must appear between the \startdata and \enddata commands
\startdata
\texttt{instrument} &  Instrument.  \\
\texttt{band} & Waveband of interest (e.g., \emph{r})  \\
\texttt{physical\_filter} &  Filter used for the exposure. \\
\texttt{day\_obs} & The observing day. \\
\texttt{group} &  Identifier for related observations. \\
\texttt{exposure} & Individual exposure. \\
\texttt{visit} &  Collection of 1 or 2 exposures. \\
\texttt{tract} &  Tessellation of the sky. \\
\texttt{patch} &  Patch within a tract.\\
\enddata

%% Include any \tablenotetext{key}{text}, \tablerefs{ref list},
%% or \tablecomments{text} between the \enddata and
%% \end{deluxetable} commands

\end{deluxetable}

A core concept of the Butler is that every dataset must be given what we call a ``data coordinate.''
The data coordinate locates the dataset in a dimensional space, where dimensions are defined in terms that scientists understand.
Some commonly used dimensions are listed in Table~\ref{tab:dims}.
Each dataset is uniquely identified by specifying its dataset type, its run collection, and its coordinates, with the Butler refusing to accept another dataset that matches all three of those values.
The dataset type defines the relevant dimensions (such as whether this is referring to observations or a sky map) and the associated Python type representing the dataset.
The run collection can be thought of as a logical grouping of datasets created by the same batch operation, but does not have to correspond to any physical structure in a file system.

As a concrete example, the file from one detector of an LSSTCam observation taken sometime in 2025 could have a data coordinate of \texttt{instrument="LSSTCam", detector=42, exposure=2025080300100} and be associated with a \texttt{raw} dataset type.
The \texttt{exposure} record itself implies other information such as the physical filter and the time of observation.
A deep coadd on a patch of sky would not have an \texttt{exposure} dimension and would instead be something like \texttt{instrument="LSSTCam", tract=105, patch=2, band="r", skymap="lsst\_cells\_v1"}, which would tell you exactly where it is located in the sky and in what waveband since you can calculate it from the tract, patch, band and skymap, where the skymap defines how tract and patch correspond to regions on the sky.

Projects can use different sets of dimension definitions, called ``universes'', and the Butler's database schema is derived from those definitions (so changing the dimension universe used by a Butler repository is possible, but not undertaken lightly, as it constitutes a schema migration).
Downstream science code generally needs to make assumptions about what is in the dimension universe (e.g., whether a \texttt{detector} concept exists and, if so, whether it \emph{implies} a \texttt{band} or can be associated with any \texttt{band}), which can limit its reusability.

\subsection{Pipelines and Tasks}

The data dimensions system also plays a fundamental role in how the LSST Science Pipelines are assembled and run; high-level pieces of algorithmic code called \texttt{PipelineTasks} declare the dimensions of their units of work (``quanta''), their inputs, and their outputs, allowing a directed acyclic graph (a ``quantum graph'') describing the processing to be assembled from a YAML declaration of the tasks to be run, their configuration, and a Butler database query.
Quantum graphs can range in size from a few tens of quanta (e.g., for the nightly processing performed on a single detector image) to millions (for a piece of the yearly data release pipelines), and serve as the common interface for multiple execution systems, including the low-latency nightly Prompt Processing framework \citep{DMTN-219} and the Batch Processing System \citep[BPS;][]{2025ASPC..538..325G}, which adapts quantum graphs for execution at scale by third-party workflow management systems like HTCondor \citep{2022zndo...2579447H}, Parsl \citep{10.1145/3307681.3325400}, and PanDA \citep{2024EPJWC.29504026K}.

Algorithmic code below the \texttt{PipelineTask} level is often subdivided into multiple ``subtasks'' that (like \texttt{PipelineTask} itself) inherit from the base \texttt{Task} class, which provides easy access to hierarchical logging, metadata, and configuration.

\subsection{Pipeline Visualization}
\label{sec:pipeline_visualization}

Visualizing pipeline execution is crucial for understanding task dependencies, debugging, optimizing workflows, and ensuring correct data flow within the LSST Science Pipelines.
To support this, we provide several options for visualizing the pipeline graph (a simplified directed acyclic graph that shows how tasks relate to dataset types but without including data from a Butler query).

\begin{figure*}
    \centering
    % \plotone{figures/middleware/pipe_viz_comcam_subset.pdf}
    % Use includegraphics to force more width for the figure.
    \includegraphics[width=\textwidth]{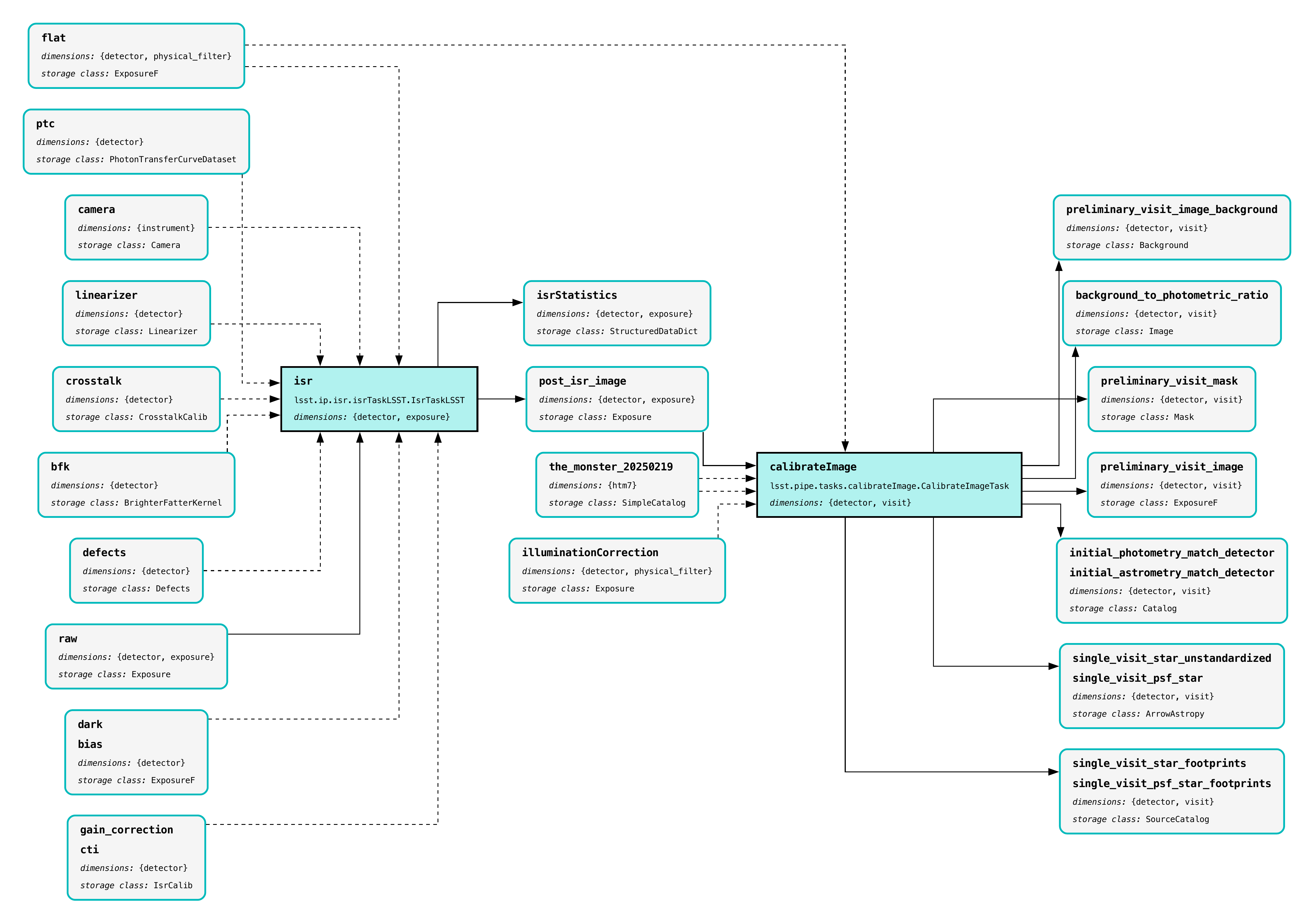}
    \caption{
        Example pipeline visualization of two selected tasks from the \texttt{LSSTComCam/DRP-v2.yaml} pipeline.
        The diagram illustrates the flow of datasets between tasks, with dashed lines indicating prerequisite calibration inputs.
        This visualization helps validate task dependencies and the expected sequence of execution.
    }
    \label{fig:pipe_viz}
\end{figure*}

Diagrams can be generated as ASCII, useful for a quick inspection from a terminal session, or in graphical formats such as Graphviz DOT\footnote{\url{https://www.graphviz.org}} or Mermaid\footnote{\url{https://mermaid.js.org}}.
The Mermaid format is particularly well-suited for sharing in accessible, web-based contexts.
Figure~\ref{fig:pipe_viz} shows a visualization of two tasks from the \texttt{LSSTComCam/DRP-v2.yaml} pipeline.
The diagram shows the relationships between tasks and their input and output dataset types as well as the sequence in which the tasks are expected to run.
Such visualizations can help uncover misconfigurations, missing inputs, or unexpected data dependencies that might otherwise result in issues such as empty quantum graphs or failed pipeline execution.
It is also possible to visualize quantum graphs, which include the actual data from a Butler query, but these can be very large and are only useful for small Butler queries to inspect the actual data flow when debugging.

\subsection{Configuration}
\label{sec:config}
The \texttt{pex\_config} package provides the foundational configuration system for running large-scale processing campaigns with the Science Pipelines.
\texttt{pex\_config} is designed to resemble a domain-specific language embedded within Python, in which configuration options are defined via Python classes  and values are set via snippets of Python code.
This leverages the full power of a programming language for parsing and setting configuration values, while also supporting interoperability with standard text formats (e.g., JSON, YAML), import-time plugin extensibility, provenance through parameter change history, and validation of both types and values.
The system can be installed and used on its own, and has been adopted by the DRAGONS software \citep{2023RNAAS...7..214L}.
A more detailed description of how the configuration system works is provided in Appendix~\ref{sec:appendix_config}.

\subsection{Instrument Abstractions: Obs Packages}
\label{sec:obs_packages}

The Butler and pipeline construction code know nothing about the specifics of a particular instrument.
In the default dimension universe there is an \texttt{instrument} dimension that includes a field containing the full name of a Python \texttt{Instrument} class.
This class, which is required to implement a specific interface, is used by the system to isolate the instrument-specific from the pipeline-generic.
Some of the responsibilities are:

\begin{itemize}
\item Register instrument-specific dimensions such as \texttt{detector}, \texttt{physical\_filter} and the default \texttt{visit\_system}.
\item Define the default \texttt{raw} dataset type and the associated dimensions.
\item Provide configuration defaults for pipeline task code that is processing data from this instrument.
\item Provide a ``formatter'' class that knows how to read raw data.
\item Define the default curated calibrations known to this instrument.
\end{itemize}

The \texttt{Instrument} interface is defined in two levels: the minimal interface in the \texttt{pipe\_base} package defines everything needed to use the Butler and execution system, while a more complete subclass in the \texttt{obs\_base} package provides considerable additional functionality but is not in the minimal, \texttt{pip}-installable suite because of the additional dependencies on C++ code.

By convention we define the instrument class and associated configuration in \texttt{obs} packages.
There are currently project-supported \texttt{obs} packages for:

\begin{itemize}
\item LSSTCam \citep{10.71929/rubin/2571927,2024SPIE13096E..1SR,2010SPIE.7735E..0JK}, LSSTComCam \citep{10.71929/rubin/2561361}, LATISS \citep{10.71929/rubin/2571930,2020SPIE11452E..0UI}, and associated Rubin Observatory test stands, calibration instruments, and simulators.
\item Hyper Suprime-Cam (HSC) on the Subaru telescope \citep{2018PASJ...70S...1M}.
\item The Dark Energy Camera (DECam) on the CTIO Blanco telescope \citep{2015AJ....150..150F,2008SPIE.7014E..0ED}.
\item Canada-France-Hawaii Telescope's MegaPrime \citep{2003SPIE.4841...72B}.
\end{itemize}

All of these instruments use the default dimension universe.

Additionally, teams outside the project have developed \texttt{obs} packages to support Subaru's `{\={O}}nohi`ula Prime Focus Spectrograph \citep{2026SPIE14149E..0CA}, VISTA's VIRCAM \citep{2015A&A...575A..25S},
the Wide Field Survey Telescope \citep[WFST;][]{2025arXiv250115018C}, Lick Observatory's Nickel Telescope \citep{2026AAS...24822403G}, and the Gravitational-wave Optical Transient Observer \citep[GOTO;][]{2021PASA...38....4M}.
SPHEREx \citep{2026ApJS..285...67A} uses the pipeline middleware but has developed pipeline tasks without the Rubin C++ dependencies and so does not use an \texttt{obs} package.
Most of these projects have made small changes to the default dimension universe to better represent their data model.

\subsection{Metadata Translation}

Every instrument uses different metadata standards but the Butler data model and pipelines require some form of standardization to determine values such as the coordinates of an observation, the observation type, or the time of observation.
To perform that standard extraction of metadata each supported instrument must provide a metadata translator class using the \texttt{astro\_metadata\_translator} infrastructure.\footnote{\url{https://astro-metadata-translator.lsst.io}}
The translator classes can understand evolving data models and allow the standardized metadata to be extracted for the lifetime of an instrument even if the headers change over time.
Furthermore, in addition to providing standardized metadata the package can also provide programmatic or per-exposure corrections to data headers prior to calculating the translated metadata.
This allows files that were written with incorrect headers to be recovered during file ingestion.

\subsection{Image Data Models and File Abstraction}

For DP2 we modified our file serialization system to make the files more approachable for external users and to improve performance when requesting subsets of a file from an object store \citep{2026SPIE14155E..17B}.
Additionally, as part of this work we changed the Python types such that a deep coadd and a visit image now have different specialized types and we are no longer trying to fit every image into a single type.
This separation of data model from serialization outputs gives us the flexibility to adopt different file formats in the future if the community begins to migrate from FITS to ASDF, for example, and will allow us to do format conversion on demand in the future (we currently support lossless conversion between FITS, HDF5, and JSON).
% <<< input{sections/middleware} <<<
% >>> input{sections/core} >>>
\section{Core Algorithmic Primitives and Data Structures}
\label{sec:core}

The high-level algorithms in the LSST Science Pipelines are largely built from algorithmic primitives and data structures implemented in the \texttt{afw} package.
\texttt{afw} is a large, complex, C++-heavy suite of multiple libraries that sometimes suffer from historical idiosyncrasies, but are nevertheless extremely powerful and well optimized.
These include (but are not limited to):
\begin{itemize}
\item \texttt{afw.geom} holds our high-level geometry primitives.  This includes composable coordinate transforms and world coordinate systems (WCS).
    It also includes \texttt{SpanSet}, a run-length encoding (RLE) description of a set of pixels in a 2-d image, a simple \texttt{Polygon} class, and routines for working with various ellipse parameterizations.
\item \texttt{afw.math} includes convolution, resampling, general-purpose interpolation, statistics, and least-squares fitting algorithms.
\item \texttt{afw.detection} contains threshold-based detection on images and point-spread function (PSF) model interfaces.
    This includes the \texttt{Footprint} class, which combines a \texttt{SpanSet} with a list of peaks to represent either a single source detection or a group of blended sources.
\item \texttt{afw.image} centers around the \texttt{Exposure} class, which combines an image (\texttt{Image}), bitmask (\texttt{Mask}), and variance image with the many objects used to astrophysically characterize an observation or coadd (PSF, WCS, aperture corrections, etc.).
      From DP2 on we have moved away from using these classes in user-facing code, preferring targeted Python types for different image types.
\item \texttt{afw.table} holds data structures for tabular data, with both row- and column-based views.  The \texttt{afw.table.io} package defines a framework for persisting arbitrary objects to a series of FITS binary table HDUs, which is used in the on-disk form of the \texttt{Exposure} class.
\item \texttt{afw.cameraGeom} provides a hierarchical description of large-format photometric cameras, such as LSSTCam, HSC, or DECam, including optical distortions, focal plane layouts, and amplifier regions.
\end{itemize}
Most of the algorithms and data structures in  \texttt{afw}  are implemented  directly within it, and often reflect the evolution of concepts originally developed in the SDSS \emph{Photo} pipeline \citep{2002AJ....123..485S} or the Pan-STARRS Image Processing Pipelines \citep{2020ApJS..251....3M}.
Others delegate to third-party libraries, particularly Eigen (linear algebra), the GNU Scientific Library (interpolation, random numbers), Boost (image iterators and geometry), CFITSIO \citep[image and table I/O;][]{2010ascl.soft10001P} and Starlink AST \citep[coordinate systems and transforms;][]{2016A&C....15...33B}.

Lower-level data structures are defined in a handful of packages just below \texttt{afw}.
The \texttt{sphgeom} package is used for spherical geometry calculations, sky-based regions, and hierarchical sky pixelization schemes, while \texttt{geom} provides simple 2-d Euclidean \texttt{Point}, \texttt{Extent}, and \texttt{Box} types in both integer- and floating-point variants.
\texttt{geom} also includes linear transforms and (for historical reasons) its own angle-manipulation and sky coordinate type.

C++ mapping types (with Python bindings) and date/time objects are defined in \texttt{daf\_base}.
The \texttt{DateTime} class is used in our C++ data structures mostly to represent TAI times.
The \texttt{PropertySet} represents a hierarchical key/value data structure whereas \texttt{PropertyList} is a flat data structure that is used to represent a FITS header and supports multi-valued keys and key comments.

Another small set of core packages sits just above \texttt{afw}:
\begin{itemize}
\item \texttt{skymap} defines interfaces and a few implementations of our system for mapping the sky onto a set of slightly overlapping image-friendly projections and tiles for coaddition.
Each distinct projection in a skymap is called a \texttt{tract}, and each \texttt{tract} is further divided into multiple \texttt{patches}.
\item \texttt{shapelet} includes optimized evaluation of Gauss--Hermite and Gauss--Laguerre functions and their derivatives.
\end{itemize}

A substantial fraction of our core packages predate the now-ubiquitous Astropy package, and in some cases we now prefer to use Astropy types in \emph{most} new code (date/time representations and tables in particular) and especially public interfaces, following the recommendations from \citet{2016SPIE.9913E..0GJ}.
Fully retiring core libraries that have Astropy counterparts is at best a long-term project, however, due to our continued need for these objects in considerable amounts of C++ that has no equivalent in Astropy (or anywhere else).
% <<< input{sections/core} <<<

\section{Key Algorithmic Components}
\label{sec:components}

Most of the key algorithms we have implemented for processing Rubin data are used in multiple pipelines, and in many cases an algorithmic component is used multiple times within a single pipeline.
For the most part, these reusable algorithms are implemented as regular \texttt{Task} objects, and these are combined in higher-level \texttt{PipelineTasks} discussed later in \secref{sec:tasks-and-pipelines}.
In other cases, the algorithm is inseparable from its I/O, and is implemented directly as a \texttt{PipelineTask}.
This section presents examples of both.

% Each of these is a subsection.
% >>> input{sections/components/isr} >>>
\subsection{Instrument Signature Removal}
\label{sec:isr}

Raw images from charge-coupled devices (CCDs) contain instrumental effects, such as dark currents, tree rings \citep{2020JATIS...6a1005P}, the brighter-fatter effect \citep{2024PASP..136d5003B}, amplifier offsets, clocking artifacts, or crosstalk between neighboring amplifiers, that can be removed in the data processing.
In the LSST Science Pipelines, this step is called Instrument Signature Removal (ISR) and is the first processing applied to a raw CCD exposure (see, for example, the pipeline workflow example shown in Fig.~\ref{fig:pipe_viz}).
The package that performs ISR on an exposure, called \texttt{ip\_isr}, is a critical component of the pipelines used to process LSST images and requires calibration products produced and verified by the \texttt{cp\_pipe} and \texttt{cp\_verify} packages, as described in \secref{sec:calib_pipe}.
% <<< input{sections/components/isr} <<<
% >>> input{sections/components/backgrounds} >>>
\subsection{Background Subtraction}
\label{sec:backgrounds}

Our low-level background subtraction algorithms (in the \texttt{meas\_algorithms} package) operate by binning images, masking out detections and bad pixels, and computing robust statistics (clipped means, by default) on the remaining values in those bins.
The bin averages can then be interpolated via Akima splines or approximated by Chebyshev polynomials.

This code is run and re-run in many different configurations in many different contexts.
For example, it is the first step run after instrument signature removal \secrefp{sec:isr} on each detector in both the nightly and data release pipelines.
In a full DRP it is also run across a full visit to remove large-scale background features \citep{2019PASJ...71..114A}.
We sometimes also run background subtraction on coadds before detecting objects on them, especially when only full-visit backgrounds were subtracted from the input images.

In addition, we routinely run background subtraction as a small tweak every time we run detection \secrefp{sec:detection}.
We \emph{define} our background to be the sum of all light that we do not consider part of a detection, which means it needs to be adjusted when the detection threshold changes (e.g., when we detect fainter objects on coadds).
% <<< input{sections/components/backgrounds} <<<
% >>> input{sections/components/detection} >>>
\subsection{Detection}
\label{sec:detection}

In the LSST Science Pipelines, \emph{detection} refers specifically to the process of finding above-threshold regions in images and one or more peaks within them (i.e., \texttt{Footprints}, as described in \secref{sec:core}).
Before this thresholding, we convolve each image with an approximation of its PSF model, as this is (approximately) optimal for detecting isolated point sources \citep{2018PASJ...70S...5B}.
After detection, \texttt{Footprints} with multiple peaks are generally deblended \secrefp{sec:deblending} before being measured \secrefp{sec:measurement}.
Our detection algorithms assume the image has already been background-subtracted \secrefp{sec:backgrounds}.

All of our detection tasks are implemented in the \texttt{meas\_algorithms} package.

\subsubsection{Source Detection}
\label{sec:SourceDetectionTask}

The \texttt{SourceDetectionTask} convolves the image with a Gaussian approximation to the exposure PSF and detects \texttt{Footprints} above a configurable threshold in either signal-to-noise or absolute flux level.
It can optionally detect negative deviations as well as positive (which is useful when operating on difference images).

A serious challenge for this algorithm in deep or otherwise crowded fields is that noise peaks or below-threshold sources can be pushed above the detection threshold when they land in the wings of brighter neighbors.
To reduce the number of spurious peaks due to this effect, \texttt{SourceDetectionTask} can optionally perform a \emph{temporary} small-scale background subtraction before looking for peaks (but after finding the above-threshold regions); while the spline and Chebyshev models are not good models for the wings of large objects, they are often better than nothing.

\subsubsection{Dynamic Detection}
\label{sec:DynamicDetectionTask}

The \texttt{DynamicDetectionTask} is a specialized version of \texttt{SourceDetectionTask} that adjusts detection thresholds based on the local background and noise.
This task was initially developed to address detection efficiency issues noted in HSC data, thought to be related to correlations in the noise due to warping.
First, the task detects sources using a lower detection threshold than normal.
In so doing, we identify regions of the sky which are unlikely to contain real source flux.
Next, a configurable number of sky objects (see \secref{sec:sky-objects}) are placed in these sky regions (1000 by default), and the PSF flux and standard deviation for each of these measurements is calculated.
Using this information, we set the detection threshold such that the standard deviation of the measurements matches the median estimated error.
More details of this algorithm can be found in \citet{2019PASJ...71..114A}.

\subsubsection{Streak Masking}
\label{sec:MaskStreaksTask}

Instead of looking for astrophysical sources like our other detection tasks, \texttt{MaskStreaksTask} is responsible for identifying and masking pixels affected by streaks (primarily artificial satellites).
It searches for linear features with a Canny filter and the Kernel-Based Hough Transform \citep{2008PatRe..41..299F}, and operates most effectively on difference images.
The algorithm is described in more detail in \citet{DMTN-197}, which focuses on its application to differences between PSF-matched warps and a PSF-matched coadd, as described in \secref{sec:coaddition-and-objects}.
This task is also used in traditional difference images, those formed by subtracting a template coadd from a science image, as described in \secref{sec:dia}.
% <<< input{sections/components/detection} <<<
% >>> input{sections/components/deblending} >>>
\subsection{Deblending}

\label{sec:deblending}

\emph{Deblending} has become the standard astronomical term for dealing with images where multiple distinct astrophysical sources overlap.
In the LSST Science Pipelines, it specifically means assigning fractions of the flux from all pixels in a \texttt{Footprint} to each of the peaks within that \texttt{Footprint}, with each peak then considered a ``child'' source.
This fractional flux assignment is very different from creating a ``segmentation map'', which assigns each pixel entirely to a single child source, as done by, for example, Source Extractor \citep{1996A&AS..117..393B}.
\texttt{HeavyFootprint} (an extension of \texttt{Footprint} that adds a flattened array of pixel values) is used to pass the per-pixel fluxes to downstream algorithms.
These \texttt{HeavyFootprint}s can also be serialized to FITS tables and are a convenient way to analyze detected sources without needing to carry the full image around.

Deblending in the Science Pipelines is performed differently for single-band (visit) image processing vs.\ multi-band (coadd) image processing.
For single-band images we use a modified version of the SDSS deblender \citep{rhldeblend} from the \texttt{meas\_deblender} package.
For multi-band images, we use a simplified version of the \textsc{Scarlet} deblending algorithm \citep{2018A&C....24..129M} from the \texttt{scarlet\_lite} package.
Our motivation for simplifying the \textsc{Scarlet} algorithm can be found in \citet{DMTN-194}.
More details on the algorithmic implementations of the deblending algorithms are given in \secref{app:deblending}.
% <<< input{sections/components/deblending} <<<
% >>> input{sections/components/measurement} >>>
\subsection{Source Measurement}
\label{sec:measurement}

After sources are \emph{detected} \secrefp{sec:detection} and optionally \emph{deblended} \secrefp{sec:deblending}, the source measurement tasks are responsible for applying a suite of measurement \emph{plugins} on the deblended pixels for each source.
Centroiders, shape measurements, and photometry algorithms are all implemented as measurement plugins.

We also distinguish between measurement on the original detection image vs. measurement on another image using that original detection.
Measurement could be performed on a single-visit image, a coadd of multiple images, or a difference of images: from the perspective of a measurement plugin, there is no difference between these cases.
\textit{Forced measurement} is performed on one image using a ``reference'' catalog of sources that were detected on another image.

The measurement tasks, plugin base classes, and a suite of standard common plugins are defined in the \texttt{meas\_base} package, as described in \secref{sec:standard-plugins}.

\subsubsection{Framework Mechanics}
\label{sec:measurement-interfaces}

Plugins are enabled or disabled in a measurement task via the task's configuration (see \secref{sec:appendix_config} for an example), and each plugin has its own configuration nested within the task configuration.
When a measurement task is constructed, it constructs instances of its enabled plugins, providing them a schema object that they can use to declare and document their output columns.
Each plugin is responsible for defining and filling in columns in the output source catalog, and almost all plugins include columns for uncertainties and at least one flag column to report failures.

Measurement plugins often depend on each other, and must be run in a particular order.
Rather than creating a directed acyclic graph to denote the dependencies, the plugins are batched and are run in any order within a batch.
The batch order is defined by the \texttt{getExecutionOrder} method, with smaller execution numbers being run first.
The base class defines a list of named constants for particular cases:
\begin{enumerate}
    \item \texttt{CENTROID\_ORDER} for plugins that require only footprints and peaks
    \item \texttt{SHAPE\_ORDER} for plugins that require a centroid to have been measured
    \item \texttt{FLUX\_ORDER} for plugins that require both a shape and centroid to have been measured
\end{enumerate}
The measurement system also provides a \textit{slot} system for predefined aliases to allow a plugin to get a value without knowing exactly what plugin originally computed that value, e.g., \texttt{slot\_Centroid} could point to \texttt{base\_SdssCentroid}, or some other plugin that measures centroids.

While the measurement tasks and plugin interfaces are Python, most concrete measurement plugins are implemented in C++, since they need to loop over pixels.

When a measurement task is run, it starts by making an empty
\texttt{SourceCatalog} (from the \texttt{afw.table} package, see \secref{sec:core}) with the plugin-defined schema and one row for each of the \texttt{Footprint} objects returned by previous detection and deblending tasks.
It then temporarily replaces all pixels within \texttt{Footprints} by random noise.
As the task loops over each row in the output catalog, that source's pixels are restored---either to the original \texttt{Exposure} pixels for isolated or otherwise un-deblended sources, or to the deblender's \texttt{HeavyFootprint} values for deblended children---and the plugins are called in execution order.
Each plugin is given the full modified \texttt{Exposure} and a row of the output catalog to fill in.
Note that plugins are \emph{not} limited to using only the pixels within a \texttt{Footprint}; they get to decide themselves which pixels to use.
After each source is measured, the task replaces its pixels with noise again, allowing the next source to be measured independently.

\subsubsection{Aperture Corrections}
\label{sec:apcorr}

With many different measures of photometry available for both stars and galaxies, establishing a consistent internal photometric system is a challenge, even before we consider the problem (covered in \secref{sec:calibration}) of mapping that system to absolute photometry via an external reference catalog.

In addition, standard PSF modeling \secrefp{sec:psf_modeling} generally focuses on the core of the true PSF, which is only adequate for photometry of galaxies and fainter stars (while the wings of the PSF matter for bright galaxies as well, the semi-arbitrary definition of the boundary of such galaxies is typically a bigger source of photometry uncertainty).

Our solution to both of these problems is \emph{aperture corrections}, in which we apply multiple photometry algorithms to a suite of isolated bright stars on each detector (by default, the same ones used to build the PSF model), compute the ratio of each algorithm to a standard one (a background-compensated top-hat aperture flux), and then interpolate that ratio using Chebyshev polynomials to other positions on the detector.
The measured fluxes of other objects are then scaled by that interpolated ratio.
This essentially forces all algorithms to produce the same results (on average) on stars.

It is clear that this approach is not exactly correct for galaxies or other extended sources in general, but we believe it is usually better than not correcting the galaxy photometry at all (it is, after all, the right thing to do for barely-resolved galaxies).

This scheme has two other serious limitations.
The first is that many of the measurements we need to correct are noisy, even on bright stars, and the interpolated ratios can add significant uncertainty to the final fluxes.
The second is that aperture corrections cannot in general be coadded along with the images (unlike PSF models); they are not a linear function of the data.
Coadding the ratios \emph{is} a valid first-order approximation in the limit where the photometry does not depend on the PSF and the PSFs of the contributing images are similar, and this is what we do at present, but we cannot currently  quantify the error this approximation introduces.
Our understanding is that using much larger PSF models is the cleanest solution to these problems, but the additional degrees of freedom that would entail may be hard to constrain with the information available.

\subsubsection{Sky Objects}
\label{sec:sky-objects}

Sky objects are a special class of object that are not detected in the image being measured, but rather placed quasi-randomly in regions that are likely to be free of real sources.
They are used to measure the sky background and noise properties of the image, and to assist in setting dynamic detection thresholds (see \secref{sec:DynamicDetectionTask}).
After these pseudo-objects are placed, they are measured and processed in the same way as a true object.
Their measurements are included in output catalogs, with an accompanying flag set to indicate that these are sky objects.

By default, sky objects are placed in regions of the sky where no detection footprints or bad pixels are located within an 8-pixel radius.
To generate initial candidate positions, we use the quasi-random Halton sequence number generator with a seed based on the image ID.
This is a deterministic sequence that mimics random sampling, but avoids clumping and gaps that can occur with true random sampling.
Doing so allows us to generate a less biased sampling of the entire sky background, often with fewer required sky objects.
The initial list of candidate positions is normally larger than the requested number of sky objects, by default of order 5 times.
Attempts are made at placing sky objects iteratively until the requested number of sky objects is reached or the candidate list is exhausted.
Typically, for visit-level detectors and coadd-level images, we place 100 sky objects.\footnote{When these pseudo-sources are placed into visit-level data, we instead refer to them as sky sources.}
Within the \texttt{DynamicDetectionTask} a larger number of temporary sky objects (default 1000) are placed in order to generate an updated detection threshold.

\subsubsection{Standard Measurement Plugins}
\label{sec:standard-plugins}

The \texttt{meas\_base} package includes a suite of standard measurement plugins.
These include second-moment shapes (\texttt{base\_SdssShape}) and adaptive centroids \citep[\texttt{base\_SdssCentroid}]{2003AJ....125.1559P}, circular aperture photometry (\texttt{base\_CircularApertureFlux}), PSF-weighted photometry (\texttt{base\_PsfFlux}), and many others.
It also includes utility plugins that transform pixel bitmask values into catalog flags (\texttt{base\_PixelFlags}) and apply coordinate transforms to centroids for use in forced photometry.
The algorithms are described in detail in \citet{2018PASJ...70S...5B}.
There are also numerous extension plugins available in other pipeline packages described in the following sections.

% >>> input{sections/components/measurement/gaap} >>>
\subsubsection{Gaussian Aperture and PSF Photometry}
\label{sec:gaap}

\texttt{meas\_extensions\_gaap} implements the Gaussian Aperture and PSF photometry (GAaP) algorithm \citep{2008A&A...482.1053K}.
It is an aperture photometry algorithm designed to obtain consistent colors of extended objects (i.e., galaxies).
This is done by weighting each (pre-seeing) region of a galaxy by the same pre-defined 2D Gaussian function in all the bands, which makes it largely insensitive to the seeing conditions in the different bands.
In practice, this is done by first convolving each object with a kernel (using the same tools described in \secref{sec:dia}) so that the PSF is Gaussian and is larger by about 15\% (this is configurable).
As a second step, each Gaussianized object is then weighted with a Gaussian aperture so that the effective pre-seeing Gaussian aperture is the same for all objects in all the bands.
The plugin is configured to use a series of circular Gaussian apertures and/or an elliptical Gaussian aperture that matches the shape of the object in the reference band.

Although the two-step approach is motivated by the original implementation in \citet{2008A&A...482.1053K}, the implementation of this algorithm within the broader context of the measurement framework makes it different from the implementation used in the Kilo-Degree Survey \citep[KiDS;][]{2024A&A...686A.170W}.
In particular, because neighboring objects are replaced with noise before measurement, Gaussianization of the PSF does not result in increased blending as mentioned in Appendix A2 of \citet{2015MNRAS.454.3500K}.
Furthermore, the uncertainty handling is different.
Correlations in noise introduced due to PSF-Gaussianization are included in the uncertainty estimates.
However, because only per-pixel noise variance is tracked, the noise treatment is forced to assume that the noise is uncorrelated to begin with, which is not true on the coadds.
See \citet{DMTN-190} for more details on the implementation.

Note that the measurements from this plugin do not produce total fluxes, but should only be used to obtain colors.
For total fluxes, measurements from CModel \secrefp{sec:cmodel} or \texttt{MultiProFit} \secrefp{sec:multiprofit} are recommended.
% <<< input{sections/components/measurement/gaap} <<<
% >>> input{sections/components/measurement/kron} >>>
\subsubsection{Kron Photometry}
\label{sec:kron}

The \texttt{meas\_extensions\_photometryKron} package implements Kron photometry \citep{1980ApJS...43..305K}.
Our Kron implementation uses a scaled version of the HSM shapes (see \secref{sec:hsm}) to form an elliptical aperture, and then scales to $2.5$ times the first radial moment.

Our implementation does not correct for the PSF in any way; this means its outputs should only be used for very well-resolved galaxies.
We do not expect our Kron photometry to be competitive with most of our other galaxy photometry algorithms in robustness or precision, but it may be useful for comparison with external measurements.
% <<< input{sections/components/measurement/kron} <<<
% >>> input{sections/components/measurement/hsm} >>>
\subsubsection{HSM Shapes}
\label{sec:hsm}

The \texttt{meas\_extensions\_shapeHSM} package contains plugins that measure the shapes of objects.

This includes a suite of PSF-uncorrected shapes (for both objects and the PSF models) and a suite of PSF-corrected shapes intended for use in weak gravitational lensing.
These are all based on the adaptive Gaussian-weighted moments algorithm described by \citet{2002AJ....123..583B} and \citet{2000AJ....120.1198F} (the \texttt{base\_SdssShape} algorithm described in \secref{sec:standard-plugins} is another implementation of this algorithm).
The implementation of these algorithms lives within the \texttt{hsm} module of the \textsc{GalSim} package \citep{2015A&C....10..121R}.

The PSF-uncorrected shapes include a variant that uses circular (rather than elliptical) Gaussian weights and another that adds noise to the PSF model image to match the noise in the data.
Both of these play important roles in measuring PSF model quality by comparing the shapes of stars to the shapes of the PSF model at the positions of those stars.

In addition to the plugins that measure (adaptive) weighted moments, there is also a series of plugins to estimate the PSF-corrected ellipticities of objects.
These are described more fully in \citet{2003MNRAS.343..459H} and \citet{2005MNRAS.361.1287M} (where ``HSM'' comes from Hirata and Seljak from the first paper and Mandelbaum from the second paper).
While no longer state-of-the-art in weak lensing, these algorithms are fast, robust, and easy to use, as long as their biases are characterized via simulations \citep{2018PASJ...70S..25M, 2022PASJ...74..421L}.

In addition to the second moments that characterize the size and ellipticity of the PSF, higher-order moments \textemdash\ those beyond second order \textemdash\ capture more subtle aspects of the PSF shape, such as skewness, kurtosis, and other asymmetric or non-Gaussian features.
The definitions of the higher moments are given in \citet{2023MNRAS.520.2328Z}.
% <<< input{sections/components/measurement/hsm} <<<
% >>> input{sections/components/measurement/trailed-sources} >>>
\subsubsection{Trailed Sources}
\label{sec:trailed-sources}

The \texttt{meas\_extensions\_trailed\-Sources} package provides a plugin that measures the properties of trailed sources, such as those caused by moving objects like asteroids or satellites.
It measures the length, angle, flux, centroid, and end points of a trailed source using the \citet{2012PASP..124.1197V} model.
This plugin is designed to refine the measurements of trail length, angle, and end points and of flux and centroid from previous measurement algorithms.
% <<< input{sections/components/measurement/trailed-sources} <<<
% >>> input{sections/components/measurement/cmodel} >>>
\subsubsection{CModel Galaxy Fitting}
\label{sec:cmodel}

The \texttt{meas\_modelfit} package's CModel algorithm is a reimplementation of the SDSS galaxy-model fitting approach, in which we fit a PSF-convolved elliptical exponential profile and de Vaucouleurs profile to each object separately, and then fit a linear combination of the two with the ellipse parameters held fixed.
This is not as principled as a true bulge-disk decomposition, in which both models are fit simultaneously, but since each fit has fewer degrees of freedom, it can nevertheless work better on low signal-to-noise or poorly resolved galaxies, where the degeneracies in a bulge-disk decomposition might otherwise lead to completely unphysical parameters.
In practice, CModel is better thought of as a crude substitute for a single-S\'ersic fit than a kind of decomposition.

Because these are PSF-convolved models, we expect CModel to provide decent photometry (including for colors) on both small and well-resolved objects.
At present, CModel flux uncertainties are known to be severely underestimated; at least some of this is due to the fact that the ellipse parameter uncertainties cannot easily be propagated into the flux uncertainties in the final fit.

The CModel code is essentially unchanged from the version used in the HSC pipelines, and additional details can be found in \citet{2018PASJ...70S...5B}.
% <<< input{sections/components/measurement/cmodel} <<<
% >>> input{sections/components/measurement/multiprofit} >>>
\subsubsection{MultiProFit Galaxy Fitting}
\label{sec:multiprofit}

MultiProFit is a package for Gaussian mixture model fitting \citep{DMTN-312}.
It is primarily used to provide multiband S\'ersic model fits to objects using all available coadds.
The \texttt{multiprofit} package and its dependencies are included in the Science Pipelines but can also be installed independently, as they only depend on packages that are also available elsewhere.

The \texttt{meas\_extensions\_multiprofit} package contains pipeline tasks (with interfaces defined in \texttt{pipe\_tasks}) necessary to run \texttt{multiprofit} on coadded and deblended images.
The first of these tasks fits a Gaussian mixture model to the PSF model image at the location of each object in a patch.
This procedure is similar to the shapelet PSF fitting functionality in \texttt{meas\_modelfit} \secrefp{sec:cmodel}.
The main differences are that the components are pure Gaussians (shapelet parameters are not supported), can have independent shapes, and are constrained to have integrals summing to unity (i.e., they are normalized).
Currently, a maximum of two components is supported; this limitation may be removed in the future.
In all cases, the structural parameters for each component are band-independent, with a separate total flux parameter for each band.
That is, individual components do not have intrinsic color gradients (although the convolved models might, if the PSF parameters vary by band).
% <<< input{sections/components/measurement/multiprofit} <<<
% <<< input{sections/components/measurement} <<<
% >>> input{sections/components/psfs} >>>
\subsection{PSF Modeling}

\label{sec:psf_modeling}

Point-spread function (PSF) modeling in the LSST Science Pipelines is largely delegated to external libraries, albeit with considerable wrapper code to adapt them to a common interface.
We use both a heavily modified version of \texttt{PSFEx} \citep{2011ASPC..442..435B,2013ascl.soft01001B} and \texttt{Piff} \citep{2021ascl.soft02024J,2021MNRAS.501.1282J} in our production pipelines.
\texttt{PSFEx} is faster and is used in nightly alert processing  as well as to obtain a preliminary model in data release processing.
\texttt{Piff} is then used in a second, more detailed round of PSF estimation during data release processing to achieve better accuracy, especially for weak gravitational lensing.

\subsubsection{PSFEx}\label{sec:meas_extensions_psfex}

The \texttt{meas\_extensions\_psfex} package provides an interface to our own library version of the \texttt{PSFEx} command-line tool.
At its core is a task that prepares selected stars for input into \texttt{PSFEx}, invokes the library, and converts the resulting model into an LSST-specific PSF object.
Key parameters such as spatial interpolation order and oversampling ratio are controlled via configuration.

For each CCD in the focal plane, \texttt{PSFEx} independently models the PSF as a linear combination of basis vectors and captures spatial variation using polynomial interpolation.
A special task offers a built-in mechanism for star selection using strict cuts on signal-to-noise ratio, FWHM range, ellipticity, and quality flags.

\subsubsection{Piff}
\label{sec:meas_extensions_piff}

The \texttt{meas\_extensions\_piff} package is a wrapper around the \texttt{Piff} package.
\texttt{Piff} is a modular package that supports various PSF models, interpolation schemes, and coordinate systems.
It can operate on a per-CCD basis or over the full field of view.
The implementation within  \texttt{meas\_extensions\_piff} does not exploit the full modularity of \texttt{Piff}; instead, it closely follows the method used for cosmic shear analysis, such as that in DES \citep{2021MNRAS.501.1282J,2025OJAp....8E..26S}; however, it has been designed to make it easy to add support for more options as needed.

The PSF model utilized is a \texttt{PixelGrid}, and the interpolation is performed using \texttt{BasisPolynomial} interpolation \citep{2021MNRAS.501.1282J}.
Modeling is executed per CCD and can employ either pixel or sky coordinates.
A key difference from \texttt{PSFEx} is that  \texttt{Piff} implements outlier rejection based on chi-squared criteria \citep[see][for more details]{2021MNRAS.501.1282J}.

Most of the options described here are adjustable through configuration.
Some important features that were implemented by \citet{2021MNRAS.501.1282J} and \citet{2025OJAp....8E..26S} have not yet been enabled but will be available in the near future.
\citet{2021MNRAS.501.1282J} configure \texttt{Piff} to fit in sky coordinates with a world-coordinate system (WCS) transform that includes CCD distortions such as tree rings.
\texttt{meas\_extensions\_piff} is capable of doing the same, but our astrometric models do not yet include CCD distortions, and hence thus far we have not used this approach in our production configuration, as it does not significantly improve the quality of the PSF models.
Additionally, although \citet{2025OJAp....8E..26S} incorporated a color correction to account for chromatic effects on the PSF, this correction has not yet been implemented in  \texttt{meas\_extensions\_piff}.
% <<< input{sections/components/psfs} <<<
% >>> input{sections/components/astro_photo_cal} >>>
\subsection{Astrometric and Photometric Calibration}
\label{sec:calibration}

Astrometric and photometric calibration in data release processing are both  performed in each of two very different steps.
First, astrometric and photometric solutions for each detector are fit independently to a reference catalog \citep{DMTN-277}, which is sufficient for matching and filtering in subsequent, more sophisticated algorithms and validation.
In the nightly alert processing, we expect to be able to use a high-density reference catalog produced by the most recent Rubin data release, and these single-detector fits represent the final calibrations.
In data release processing, much more sophisticated final astrometric and photometric transforms are then fit to catalogs from multiple epochs at once.

The LSST Science Pipelines use the Starlink AST library \citep{2016A&C....15...33B} for persisting, composing, and evaluating coordinate transformations.
We have our own class in the \texttt{afw} package for representing (among other things) photometric calibrations, usually with Chebyshev polynomials.
These can also be multiplied and even mixed with AST-backed transform objects to represent pixel-area corrections.
Our coordinate transforms are not exactly representable via the FITS WCS standard \citep{2002A&A...395.1061G}, which has no way to describe a chain of composed transforms.
For public data releases, we approximate these transforms via a FITS TAN-SIP WCS \citep{2005ASPC..347..491S} for convenience, but precision astrometry should always use the exact composed transformation.

\subsubsection{Single-Frame Astrometric Calibration}
\label{sec:astrometryTask}

Single-frame astrometric fits are performed by a task in the \texttt{meas\_astrom} package.
This task matches a catalog of sources detected and measured on an image to a reference catalog and solves for the \textit{World Coordinate System} (WCS) of the image.
Matching and WCS fitting are performed iteratively, to reject astrometric outliers.
The matcher is either the optimistic or pessimistic matcher from \citet{2007PASA...24..189T}, with the pessimistic matcher used by default due to better performance on dense fields; see \citet{DMTN-031} for details.
The WCS fitter can be a simple affine model on top of fixed camera geometry or a FITS TAN-SIP WCS.
We default to fitting the simple affine model because we have good distortion models for most of the instruments we support, often from previous fitting with \texttt{gbdes} \secrefp{sec:gbdes}, and hence we do not need the extra degrees of freedom provided by a TAN-SIP model.

\subsubsection{GBDES}
\label{sec:gbdes}

The final astrometric solution in DRP is fit by a task in \texttt{drp\_tasks}, which runs the \texttt{wcsfit} fitter from the \texttt{gbdes} package \citep{2022ascl.soft10011B,2017PASP..129g4503B} on the ensemble of images in a given band overlapping with a given tract.
This task fits a per-detector polynomial distortion model, a per-exposure polynomial distortion model, and positions for all the isolated star sources in the component images.
This is done by first associating all isolated point sources in the input images and matching them with an external reference catalog.
The model is then fit by iterating between fitting the per-detector and per-exposure polynomial models, and recalculating the best-fit solution for the object positions.

The task can be configured to fit either a two-parameter (position on the sky) or five-parameter (position, proper motion, and parallax) solution for the input objects.
Correcting for differential chromatic refraction is another configurable option.

There are also options to run variants of the main task: one that fits images from multiple bands at once, in which case the per-detector distortion model is also per-band; one that removes the restriction to a single tract and fits images regardless of their location on the sky by splitting the images into contiguous groups; and one that combines these two options.

Lastly, the per-detector polynomial model fit by the task is also used to build a camera distortion model, which can be fed back into single-frame modeling or into the \texttt{gbdes} fit for other data.
For use in single-frame modeling, a subtask is used to build an \texttt{afw} \texttt{Camera} object out of the native polynomial model.

Once the astrometric model has been calculated, it is exported into a form understood by the AST library.
A full description of astrometric calibration in the pipeline is given in \citet{DMTN-266}.

\subsubsection{Single-Frame Photometric Calibration}
\label{sec:photoCal}

Single-frame photometric calibration is performed by a task in the \texttt{pipe\_tasks} package.
The catalog to be calibrated is down-selected to contain only bright ($S/N>10$), well-measured, PSF-like sources which are then matched to a reference catalog.
The matched sources have their instrumental fluxes converted into rough magnitudes, which are iteratively compared with the reference catalog magnitudes using a sigma-clipping algorithm, to fit a single magnitude zero point to the whole image.
Precomputed color terms can also be applied to the reference catalog fluxes when needed.

\subsubsection{FGCM}
\label{sec:fgcmcal}

Global photometric calibration for DRP is computed via the Forward Global Calibration Method~\citep[FGCM;][]{2018AJ....155...41B}, as adapted for LSST \citep{SITCOMTN-086}.
This global calibration algorithm makes use of repeated observations of stars in all $ugrizy$ bands, combining a forward model of the atmospheric parameters with instrumental throughputs measured with auxiliary information.
In this way we simultaneously constrain the atmospheric model as well as standardized top-of-atmosphere (TOA) star fluxes over a wide range of star colors, including full chromatic corrections from the instrument and atmosphere.
The software is designed to be able to track changes in telescope components over time such as the primary mirror emissivity.

Running \texttt{fgcmcal} first requires generating a look-up table.
The input to the look-up table includes the effect of a MODTRAN \citep{1999SPIE.3756..348B} atmospheric model at the elevation of the observatory, as well as the throughput as a function of wavelength and position from the optics, filters, and detector quantum efficiency.
The quality of the output (in terms of repeatability of bright isolated stars across a wide range of colors) depends on the knowledge of the instrumental throughput.

The primary goal of \texttt{fgcmcal} is to provide a uniform relative photometric calibration of the survey.
For ``absolute'' (relative) calibration, a reference catalog can be used as an additional constraint on the fit.
Thus, the overall throughput output by \texttt{fgcmcal} depends on the reference catalog.
This can be checked with, for example, specific white dwarfs or CALSPEC \citep{2007ASPC..364..315B} stars in the survey.
However, the relative spatial and chromatic calibration of the \texttt{fgcmcal} calibration means that the absolute calibration reduces to a set of 6 numbers (one for each band, or one overall throughput and 5 absolute colors).

\subsubsection{jointcal}
\label{sec:jointcal}

The \texttt{jointcal} package provides an algorithm that fits both astrometry and photometry across multiple exposures of large mosaic cameras, fitting for both the true star positions/fluxes, and the distortions caused by the telescope and instrument.
\texttt{jointcal} is no longer used by the LSST camera pipeline, but is available for use by cameras that are not supported by \texttt{gbdes} and/or \texttt{fgcmcal} (for example, DECam).
More details on the \texttt{jointcal} algorithm are available in \citet{DMTN-036}.
% <<< input{sections/components/astro_photo_cal} <<<
% >>> input{sections/components/schemas} >>>
\subsection{Catalog Schemas}
\label{sec:schemas}

Pipeline products must be transformed from the internal data model to the public data model defined in \citet{LSE-163}.
A set of YAML files in the \texttt{pipe\_tasks} repository are used for transforming the internal pipelines representation of the data to a standardized Parquet output format.
These Parquet files are continually validated against an appropriate schema to ensure that the column names and types are correct as the pipelines codebase evolves.
For data previews and releases, the Parquet files are then ingested into the Qserv database \citep{2025ASPC..538..114M}, where the data catalogs are stored.

The public data model is defined by a set of files in the Felis \citep{2024arXiv241209721M} YAML format which describe the schema of a data catalog, including its tables, columns, constraints and metadata.
These are collectively referred to as the Science Data Model (SDM) schemas.
The YAML files are managed via the \texttt{sdm\_schemas} package with all changes validated by GitHub workflows.
The schemas corresponding to Science Pipelines output are continually evolving with the pipelines codebase, so, for instance, column names and types may be updated to reflect changes to the internal data model.
Schemas for data previews and releases represent a snapshot of the public data model at the time of the release and would typically only be updated with bug fixes, minor changes, or updates and additions to the metadata.
For public-facing data catalogs, the Felis representation is used to generate a \texttt{TAP\_SCHEMA} database describing the tables and columns available in the TAP service \citep{2019ivoa.spec.0927D} and serves as a source of documentation.
% <<< input{sections/components/schemas} <<<

\section{High-level Tasks and Pipelines}
\label{sec:tasks-and-pipelines}

In this section, we describe how the reusable components of \secref{sec:components} are assembled into pipelines for generating nightly alerts immediately from just-observed data, producing cumulative data releases roughly every year, and creating calibrations that are input to both on many cadences in between.
This also includes some description of algorithmic details that are specific to certain pipelines.

% Each of these is a subsection.
% >>> input{sections/tasks-and-pipelines/single-frame} >>>
\subsection{Single-Frame Processing and Calibration}
\label{sec:sfp}

\texttt{CalibrateImageTask}, from the \texttt{pipe\_tasks} package, performs ``single frame processing'' on a post-ISR \secrefp{sec:isr} single-detector exposure.
We repair and mask cosmic rays and defects, perform an initial set of detection \secrefp{sec:detection} and measurement \secrefp{sec:measurement} passes to estimate the image PSF, compute an astrometric \secrefp{sec:astrometryTask} and photometric \secrefp{sec:photoCal} calibration, and compute summary statistics on the resulting exposure and catalog.
The primary user-facing outputs of this task are the photometrically calibrated, background-subtracted \texttt{preliminary\_\-visit\_\-image}, the calibrated \texttt{preliminary\_\-visit\_\-image\_\-background} that was subtracted from it, and \texttt{single\_\-visit\_\-star\_\-unstandardized}, a catalog of bright point-like sources that were used as inputs to calibration, with only a small number of measurements performed on them.
As this task only processes a single detector, it is used by both DRP \secrefp{sec:drp_pipe} and AP \secrefp{sec:ap_pipe}, though with different configurations (AP is focused on latency, while DRP performs more measurements).

In DRP, \texttt{ReprocessVisitImageTask}, from the \texttt{drp\_tasks} package, takes the outputs of the global astrometric and photometric models, a visit-level background and PSF model, and re-runs detection and measurement on the post-ISR single-detector exposure.
The primary user-facing outputs of this task are the photometrically calibrated, background-subtracted \texttt{visit\_image}, the calibrated \texttt{visit\_image\_background} that was subtracted from it, and \texttt{source\_unstandardized}, a catalog of all sources detected to $5\sigma$, with all relevant measurements performed on them (this is later schema-standardized and consolidated into the \texttt{source} per-visit catalog).
In addition to  its primary purpose of performing detection and measurement using the ``best'' available inputs, this task also allows us to save only a small number of relatively small-sized intermediate products in order to regenerate the  \texttt{visit\_image} from a \texttt{raw} image, thereby reducing long-term storage needs.
% <<< input{sections/tasks-and-pipelines/single-frame} <<<
% >>> input{sections/tasks-and-pipelines/coaddition-and-objects} >>>
\subsection{Coaddition and Object Tables}
\label{sec:coaddition-and-objects}

Coadded images are used both as static-sky templates for image subtraction and for detecting and measuring faint objects.
The coaddition process is divided into two main stages: resampling the input images onto a common projection (``warping'') defined by a \texttt{skymap} \secrefp{sec:core} and stacking those resampled images into a single coadd.
By default we choose a sky map where the pixel scale matches the nominal detector pixel scale, although the sky map definition is fully configurable.
Each stage is implemented via configurable tasks that allow the pipelines to be adapted for different instruments and observing strategies.
An optional intermediate step can also be used to convolve the warped images to a configurable, common model PSF.
This PSF-homogenized variant is useful when the scientific goals require uniform PSF properties across the coadd and is used for artifact rejection during coaddition.

All warping tasks use a configurable interpolation kernel.
A 5th-order Lanczos kernel is used by default, balancing fidelity and computational efficiency while also ensuring we conserve flux by normalizing the warping kernel by its per-output-pixel sum, with a nearest-neighbor kernel for the integer bit mask.
Input images are geometrically transformed using the WCS and interpolated onto the target projection defined by a tract and patch geometry.
The input WCS includes camera distortions, at the level of about 0.2\% across the focal plane \citep{LCA-11591}, as well as corrections for atmospheric distortions \citep{DMTN-324}.
Each resulting resampled image is called a \emph{warp}.

Once warps are generated, they are stacked into a final coadd.
The default implementation performs outlier rejection to remove transient artifacts such as cosmic rays, ghosts, satellite trails, and moving objects.
The algorithm compares pixel values across epochs and masks those that significantly deviate from the expected distribution.
The artifact rejection algorithm is detailed in \citet{DMTN-080}.
By default, weights for stacking are derived from the inverse of the average variance of each warp, with optional filters on PSF quality and seeing.
The stacked image is accompanied by a mask plane and variance map, and the set of input PSF models is combined into a spatially varying coadd PSF model (\texttt{CoaddPsf}) to serve as the PSF model for the coadd.

Deep object catalogs are generated from coadds by running our detection code (see \secref{sec:detection}) on the coadd from each band independently, and then spatially matching the detected peaks across bands.
The combined cross-band \texttt{Footprint} objects are formed from the union of the per-band ones.
These are then passed to our multi-band deblender (see \secref{sec:multiband_deblending}).

While a few of our measurement codes can then run natively on multiple bands together (e.g., the galaxy fitting codes described in \secref{sec:multiprofit}), for most algorithms we obtain consistent cross-band measurements in two phases.
First, we measure on each band independently (albeit using the multi-band deblender outputs); then we measure again in each band in forced photometry mode, holding the position and (when relevant) shape fixed to the values measured in that object's reference band.
% <<< input{sections/tasks-and-pipelines/coaddition-and-objects} <<<
% >>> input{sections/tasks-and-pipelines/dia} >>>
\subsection{Difference Image Analysis}
\label{sec:dia}

Image subtraction for transient/variable detection and analysis is implemented in the \texttt{ip\_diffim} package, and is divided into three steps.
While the image subtraction system could be used to work on other combinations, we primarily focus on subtracting a ``template'' coadd from a single-visit ``science'' image.
First, the template coadd is warped to the WCS and bounding box of the science image.
Then the warped template is subtracted from the science image using the standard Alard \& Lupton algorithm \citep{1998ApJ...503..325A}, which produces a temporary difference image.
Finally, peaks are detected on the difference image and DIASources (``difference image analysis sources'') are measured.
The final difference image with updated mask planes is written along with the DIASource catalog.

\subsubsection{PSF Matching and Subtraction}

The primary implementation of image subtraction is based on \citet{1998ApJ...503..325A}, and uses spatially varying Gauss--Hermite basis functions for the fit.
The PSF-matching kernel can be constructed for either the science or the template image, and the resulting difference image is decorrelated \citep{DMTN-021}.
Optionally, the science image can be preconvolved with its own PSF before PSF-matching, producing a Score image analogous to \citet{2016ApJ...830...27Z}.

\subsubsection{DIA Detection and Measurement}
\label{sec:detectAndMeasureDiaSource}

Positive and negative peaks are detected by thresholding the Score image if it is available.
Otherwise, the difference image is smoothed with a Gaussian of the same width as the PSF of the science image, and thresholds are taken on the smoothed image.
Contiguous pixels around each peak that are statistically brighter than the background are grouped into source footprints, and any overlapping footprints are merged.
Footprints that contain both a positive and a negative peak are fit as dipoles.
The dipole fit simultaneously solves for the negative and positive lobe centroids and fluxes using non-linear least-squares minimization.
DIASources that are not classified as dipoles instead fall back on the \texttt{base\_SdssCentroid} \secrefp{sec:standard-plugins}.
Finally, all configured measurement plugins are run, including HSM shape measurements \secrefp{sec:hsm} and a trailed-source fit \secrefp{sec:trailed-sources}.

\subsubsection{Filtering Non-astrophysical DIASources}
\label{sec:streaks}

A subset of the DIASources detected during \texttt{detectAndMeasureDiaSource} are expected to be non-astrophysical.
These may be due to a variety of causes, including uncorrected instrument signatures, diffraction spikes and wings of bright stars, unmasked cosmic rays, algorithmic failures in background subtraction or image differencing, and satellites or space debris orbiting the Earth.
The latter can leave long streaks that cross one or more LSSTCam detectors
\citep{2020AJ....160..226T,2022A&C....3900584H}.

To reduce the number of DIASources originating from non-astrophysical sources, we apply a multipronged approach:

\begin{itemize}
\item During detection, we remove sources that have pixel flags characteristic of artifacts and unusable measurements, such as when all of the central pixels of a DIASource are saturated.

\item In AP, we identify and filter out DIASources that are spatially and temporally coincident with the predicted positions of cataloged artificial satellites.

\item DIASources with negative forced fluxes on the direct science image characteristic of bad background subtraction are filtered.
      This and subsequent catalog filtering is carried out by a task in the \texttt{ap\_association} package.

\item Trailed DIASources consistent with a rate of motion greater than 10 degrees per day are filtered.

\item Blank sky sources \secrefp{sec:sky-objects} used for noise estimation are removed from the output catalog.

\end{itemize}

No pixels are altered or redacted during the steps above.
The machine-learned reliability score \secrefp{sec:reliability} provides a further diagnostic that users can apply to remove remaining artifacts.
These approaches are intended to enable science users to control the tradeoff between completeness and purity during their analysis while ensuring that the pipelines and databases can maintain their required performance.

% >>> input{sections/tasks-and-pipelines/reliability} >>>
\subsubsection{Reliability Scoring}
\label{sec:reliability}

The \texttt{meas\_transiNet} package determines a numerical score for input cutout images using pre-trained machine-learning models.
Image differencing may produce false detections, so time-domain surveys characteristically use machine-learning classifiers to distinguish astrophysical sources from artifacts \citep[``Real/Bogus;'' e.g.,][]{2012PASP..124.1175B, 2015AJ....150...82G, 2019MNRAS.489.3582D}.

The \texttt{meas\_transiNet} package defines ``model packages'' that consist of a Python architecture class, a PyTorch \citep{NEURIPS2019_bdbca288} weights file, and associated metadata.
The inference task may be configured to load a model package from disk or from the Butler.

The \texttt{RBTransiNetTask} PipelineTask takes as input three square cutouts of configurable size from the science, template, and difference images centered on the location of a source.
These images are concatenated, batched into Torch blobs, and passed to the model for inference.
Either CPU or GPU backends may be used for inference.
The output of the task is a single float ranging from 0 to 1 for each cutout triplet, with higher values indicating that the DIASource is more likely to be astrophysical.
These reliability scores are then joined with the DIASource catalogs by a later transformation task.
Detailed discussion of the model architecture, training, and performance will be presented in T.\ Acero-Cuellar et al.\ (in prep.).
% <<< input{sections/tasks-and-pipelines/reliability} <<<

\subsubsection{Source Association}
\label{sec:association}

The \texttt{ap\_association} package contains multiple tasks for standardizing newly detected DIASources and associating them into ``DIAObjects''.
Standardization converts the output catalogs from difference imaging to the format specified in \texttt{sdm\_schemas} \secrefp{sec:schemas}, and applies filtering consistent with \citet{DMTN-199}.
Once DIASource catalogs are standardized, they are associated with DIAObjects in either of two modes: DRP or AP.
Both implementations use the Pessimistic Pattern Matcher B \citep{DMTN-031} to score and match DIASources, but differ in how DIAObjects are stored and how visits are ordered.

\begin{itemize}
\item DRP association loads all DIASource catalogs from a set time period overlapping a single patch at once, and creates new DIAObjects for matched DIASources from all visits simultaneously.
\item AP association processes a single visit at a time, and creates new DIAObjects incrementally from unassociated DIASources.
DIAObjects and their associated DIASources are stored in the Alert Production Database (APDB; \secref{sec:apdb}).
\end{itemize}

After association, an additional filtering step may be applied to DIASources with no matched DIAObject or Solar System object \secrefp{sec:solsys}.
Properties of the source such as its reliability score \secrefp{sec:reliability}, source flags, or signal-to-noise cuts may be used to drop detections that are likely to be false detections and avoid creating erroneous new DIAObjects.

\subsubsection{Alert Production Database (APDB)}
\label{sec:apdb}

The Alert Production Database \citep[APDB;][]{DMTN-293} supports SQLite, PostgreSQL, and Apache Cassandra database formats.
SQLite and PostgreSQL are used for testing only; a more scalable Cassandra database is used in production.
The previous history of DIAObjects, DIASources, and DIAForcedSources for the region containing the science image is loaded and passed to a task for association.
Loading is split from the association step to enable preloading of catalogs from the database in Prompt Processing during the interval when the next visit has been scheduled but the images have not yet been taken.
When AP-style association is run outside of Prompt Processing, it is therefore essential to process all association tasks in strict visit order to prevent loading catalogs from the APDB prematurely and losing DIAObject history in association.

\subsubsection{Alert Generation}
\label{sec:alerts}

In order to enable real-time science, the AP pipelines generate alert packets for each detected DIASource.
These packets are serialized in Apache Avro\footnote{\url{https://avro.apache.org/}} format and then transmitted to community alert brokers via Kafka for further processing.
\citet{DMTN-093} provides a high-level overview of the alert system.

Within the pipelines, alert packets are constructed within and published by \texttt{ap\_association}.
Alert packets contain the triggering DIASource record; the associated DIAObject or SSObject record; up to twelve months of past history from DIASources, DIAForcedSources, and/or upper limits; and cutout images of the science, template, and difference images centered at the position of the DIASource.
Cutouts are provided as FITS images serialized by the Astropy \texttt{CCDData} class, and include image, variance, and mask planes along with WCS information and an image of the approximate PSF.

Avro schemas are stored in the \texttt{alert\_packet} package.
They are derived from the corresponding AP schemas in \texttt{sdm\_schemas} used to instantiate the AP databases.
% <<< input{sections/tasks-and-pipelines/dia} <<<
% >>> input{sections/tasks-and-pipelines/solsys} >>>
\subsection{Solar System Pipelines}
\label{sec:solsys}

\begin{figure*}
\begin{center}
\plotone{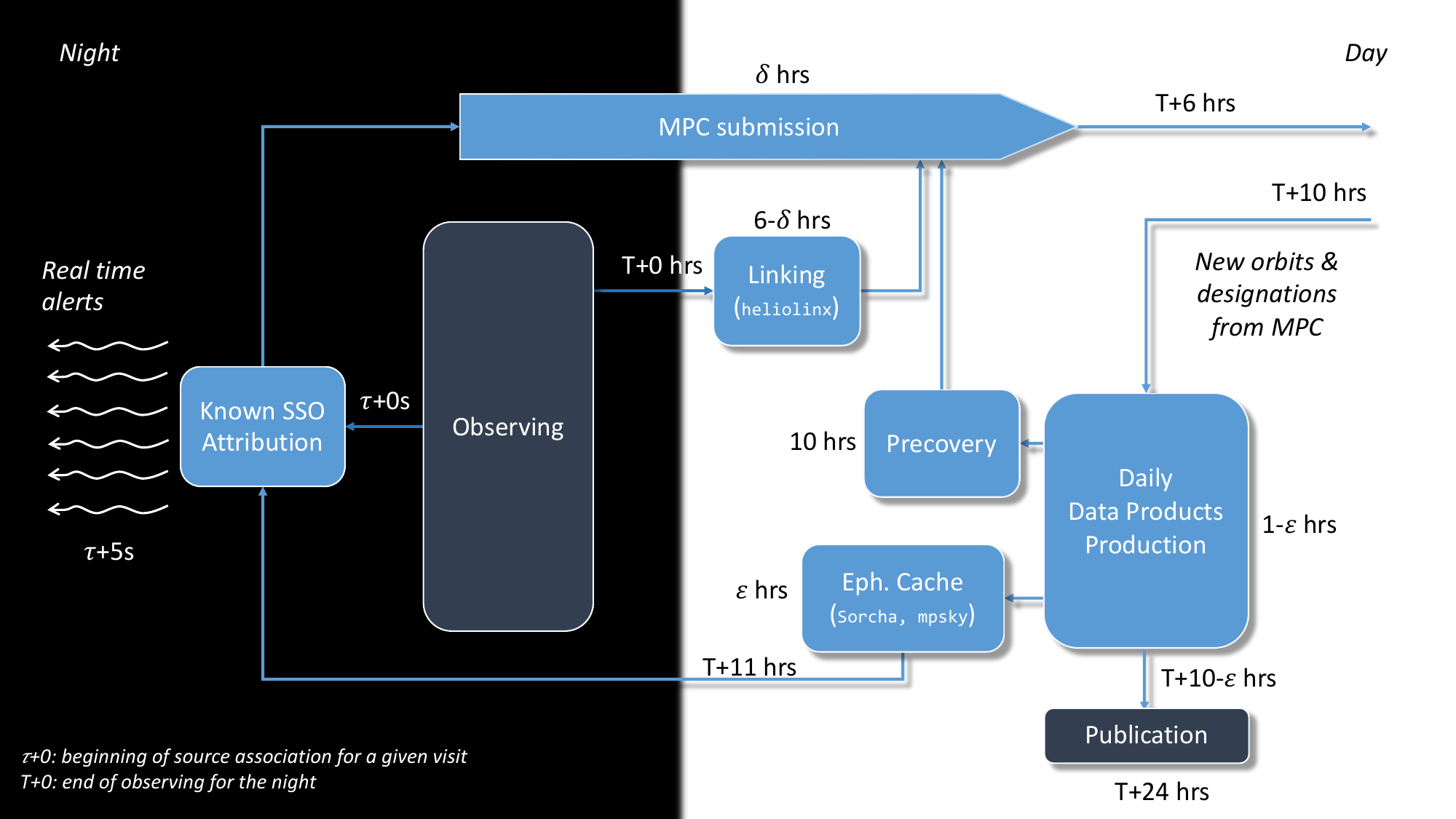}

\caption{\label{fig:ssp} Detection, attribution, linking, submission and precovery of moving sources within the nightly data: The attribution is performed in real time by the AP pipelines querying the \texttt{mpsky} service with resulting information attached to the alerts and queued for submission to the MPC.
The linking is performed in daytime using \texttt{heliolinx}, with resulting links queued for submission to the MPC.
Fetching of data from the MPC is performed automatically using PostgreSQL replication, with new data triggering recomputation of physical properties and precovery runs in the Daily Data Products Pipeline.
Any observations discovered by the precovery procedure are queued for submission to the MPC, using the submission manager tool.
The ephemeris cache is precomputed at dusk using \texttt{Sorcha} and \texttt{mpsky}, to enable fast attribution at nighttime.
All timings denote design goals.
}

\end{center}
\end{figure*}
The Solar System Pipeline (SSP; Figure~\ref{fig:ssp}) suite is responsible for (i) discovering previously unknown Solar System objects by linking together observations (usually DIASources) unattributable to static (non-moving) objects, (ii) reporting these to the Minor Planet Center (MPC), (iii) computing basic physical characteristics such as absolute magnitudes and slope parameters for all asteroids where sufficient data is available, and (iv) using the orbits received from the MPC to associate their apparitions in the DIASource tables (both in real time and as precovery in DRP).
Unlike the other packages described in this paper, not all of the Solar System Pipeline is installed as part of the standard Science Pipelines software via the \texttt{lsst\_distrib} metapackage.

The core element of the SSP is the linking pipeline, named \texttt{heliolinx} \citep{heliolinx}.
This code, run in daytime, clusters newly detected DIASources to search for candidate asteroids.
The high-level procedure is to link DIASource detections within a night (when on-sky motion is approximately linear) into \emph{tracklets}, to link these tracklets across multiple nights (into linkages) and to fit the linkages with an orbital model to identify those linkages that are consistent with an asteroid orbit.
The Rubin implementation of this software (Heinze et al., in prep.) is based on the HelioLinC algorithm \citep{2018AJ....156..135H}, with the key change being that the clustering is performed not on the sky, but in 3D space.
It is designed to be capable of detecting 95\% of all Solar System objects whose tracklets are observed over three nights within a 15-night window.\footnote{Detailed criteria are specified in the LSST Observatory System Specification (OSS) document \texttt{OSS-REQ-0159} \citep{LSE-30}.}
\texttt{heliolinx} is written in C++, but provides a Python API via \texttt{solsys\_pipe}.

Candidate discoveries that are highly certain to be real, as well as re-observations of already known objects, are reported to the Minor Planet Center (MPC) using the observation submission pipeline.
The astrometric and photometric data are converted to the PSV variant of the Astrometric Data Exchange Standard \citep[ADES;][]{2017DPS....4911214C}, and submitted to the MPC.

Following processing and validation, newly reported candidates are added to the MPC's central database.
This database, including the table of orbits as well as observations, is replicated to the US Data Facility using PostgreSQL logical replication.
Following the replication, the absolute magnitudes of objects in the SSObject table will be recomputed, as well as some auxiliary per-observation information for individual observations (the SSSource table).

The replicated orbits and computed absolute magnitudes are utilized to predict positions (ephemerides) and magnitudes of Solar System objects in subsequent nights.
To enable speedy retrieval (of order 100\,ms or less) of all objects in a visit, we precompute on-sky locations of all Solar System objects at twenty times spread across the night, fit Chebyshev polynomials, and build an efficient HEALPix-based index allowing for fast lookup.
These ephemerides are then served to association pipelines described in \secref{sec:association}.
This element of the pipeline is based on Sorcha \citep[computation;][]{2025AJ....170..100M} and \texttt{mpsky} \citep[fast lookup and serving;][]{mpsky}.
The system can also perform ``precovery''---the association of originally missed observations of Solar System objects observed earlier in the survey.

Taken together, this suite of pipelines enables Rubin to identify sources consistent with being observations of new and previously known objects in the Solar System, and makes these data public by reporting their discoveries to the Minor Planet Center and making them available to Rubin users via the Prompt Products Database \citep[PPDB;][]{DMTN-293}.
% <<< input{sections/tasks-and-pipelines/solsys} <<<
% >>> input{sections/tasks-and-pipelines/cp_pipe} >>>
\subsection{Calibration Pipelines}
\label{sec:calib_pipe}

The high-level pipelines to build calibration products (\texttt{cp}) for the LSST cameras are defined in \texttt{cp\_pipe}.
They set ISR (see \secref{sec:isr}) configuration parameters needed for each calibration product, consisting of all the sequential steps of the ISR task up to the step before the correction being generated.
In some cases, configurations also specify whether to combine exposures (for bias or dark exposures for instance) and to bin exposures to support diagnostic displays.
In LSST cameras, calibration products typically include a combined bias, a combined dark, a Photon Transfer Curve (PTC), a crosstalk matrix, a list of defects, and a look-up table of non-linearity parameters.

Once calibration products are produced, they are ``verified'' (see \citet{DMTN-222} and \citet{SITCOMTN-086} for more details) using \texttt{cp\_verify} pipelines by checking that they pass metrics defined in \citet{DMTN-101}.
In this case, verify configuration parameters enable all corrections in the ISR task up to and including the application of the correction being verified.
As a result, the calibration products can then be certified to be available in the Butler and used for ISR of an exposure.
A general overview of the ISR steps and calibration products production (including generation, verification, certification, approval, and distribution) is given in \citet{2025JATIS..11a1209P}.
% <<< input{sections/tasks-and-pipelines/cp_pipe} <<<
% >>> input{sections/tasks-and-pipelines/ap_pipe} >>>
\subsection{\texttt{ap\_pipe}}
\label{sec:ap_pipe}

The \texttt{ap\_pipe} package defines the pipeline(s) to be used for real-time Alert Production processing.
These pipelines include instrument signature removal \secrefp{sec:isr}, calibration \secrefp{sec:calibration}, measurement plugins \secrefp{sec:measurement}, image differencing \secrefp{sec:dia}, source association \secrefp{sec:association}, and alert generation \secrefp{sec:alerts}.
Some of these tasks are shared with the pipelines in \texttt{drp\_pipe} \secrefp{sec:drp_pipe}, but are configured to prioritize speed over strict quality; for example, they use a minimal set of measurement plugins.

\texttt{ap\_pipe} currently has pipeline variants for LSSTCam, LSSTComCam, LATISS, the Rubin Observatory simulators, Hyper Suprime-Cam, and the Dark Energy Camera.
Because these variants serve as testbeds for AP-specific algorithms and configuration settings, they are, as much as possible, the ``same'' pipeline, differing almost entirely in loading instrument defaults from \texttt{obs} packages \secrefp{sec:obs_packages}.
The only other customization is an extra task for handling DECam's inter-chip crosstalk, which does not have an equivalent for Rubin instruments.
% <<< input{sections/tasks-and-pipelines/ap_pipe} <<<
% >>> input{sections/tasks-and-pipelines/drp_pipe} >>>
\subsection{\texttt{drp\_pipe}}
\label{sec:drp_pipe}

The \texttt{drp\_pipe} package defines the pipeline(s) to be used for the annual Data Release Production processing.
These pipelines include instrument signature removal \secrefp{sec:isr}, calibration \secrefp{sec:calibration}, measurement plugins \secrefp{sec:measurement}, coaddition and coadd processing \secrefp{sec:coaddition-and-objects}, image differencing \secrefp{sec:dia}, source association \secrefp{sec:association}, and global calibration \secrefp{sec:calibration}.
Some of these tasks are shared with the pipelines in \texttt{ap\_pipe}, but are configured to prioritize accuracy over speed.

\texttt{drp\_pipe} currently has pipeline variants for LSSTCam, LSSTComCam, LATISS, the Rubin Observatory simulators, Hyper Suprime-Cam, and the Dark Energy Camera.
Because these variants serve as testbeds for DRP-specific algorithms and configuration settings, they are, as much as possible, the ``same'' pipeline, differing primarily in loading instrument defaults from \texttt{obs} packages \secrefp{sec:obs_packages}.
The pipelines for some instruments also disable certain multi-visit calibration routines like FGCM \secrefp{sec:fgcmcal} and GBDES \secrefp{sec:gbdes}, as those can take significant effort to commission on a new instrument, and this has not been done in all cases.
The only other customization is an extra task for handling DECam's inter-chip crosstalk, which does not have an equivalent for Rubin instruments.
% <<< input{sections/tasks-and-pipelines/drp_pipe} <<<

\section{Analysis Tooling}
\label{sec:analysis}

% Each of these is a subsection.
% >>> input{sections/analysis/display} >>>
\subsection{Display Abstractions}
\label{sec:display}

The \texttt{afw.display} subpackage defines a simple image display abstraction layer that can be used to show our image objects via multiple applications and libraries.
This includes programmatic control of stretch levels and colors, WCS mapping, semi-transparent bitmask overlays, and simple geometric region support.

There are currently implementations for matplotlib \citep{2007CSE.....9...90H}, Firefly \citep{2020ASPC..527..243R}, SAOImage DS9 \citep{2003ASPC..295..489J}, and Ginga \citep[][via Astrowidgets]{2013ASPC..475..319J}.
While most of these tools have considerable functionality beyond what our abstraction layer provides, the ability to interact easily with them in a consistent, programmatic way in many different contexts (e.g., DS9 on personal machines, matplotlib and Ginga in Jupyter notebooks, and Firefly in the Rubin Science Platform) is invaluable, and usually the minimal \texttt{afw.display} interface provides all we need.
% <<< input{sections/analysis/display} <<<
% >>> input{sections/analysis/analysis_tools} >>>
\subsection{Analysis Tools}

The \texttt{analysis\_tools} package provides a framework to allow reproducible, automatic creation of plots and metrics through a set of configurable, reusable tools that can be used in pipeline execution and interactive analysis.
The package allows metrics and plots to be consistently created at various points in the pipeline and ensures that the metrics dispatched to the monitoring dashboard are generated in sync with the archived plots.

\texttt{analysis\_tools} is built on top of \texttt{pex\_config} (see \secref{sec:config} for details); this allows it to integrate into the rest of the software pipelines and to be very flexible in its application.
It also means that the configurations for all tools are saved every time, allowing them to be accurately recreated with information such as the signal-to-noise ratio used to filter the data.

\begin{figure}[ht]
\includegraphics[width=\linewidth]{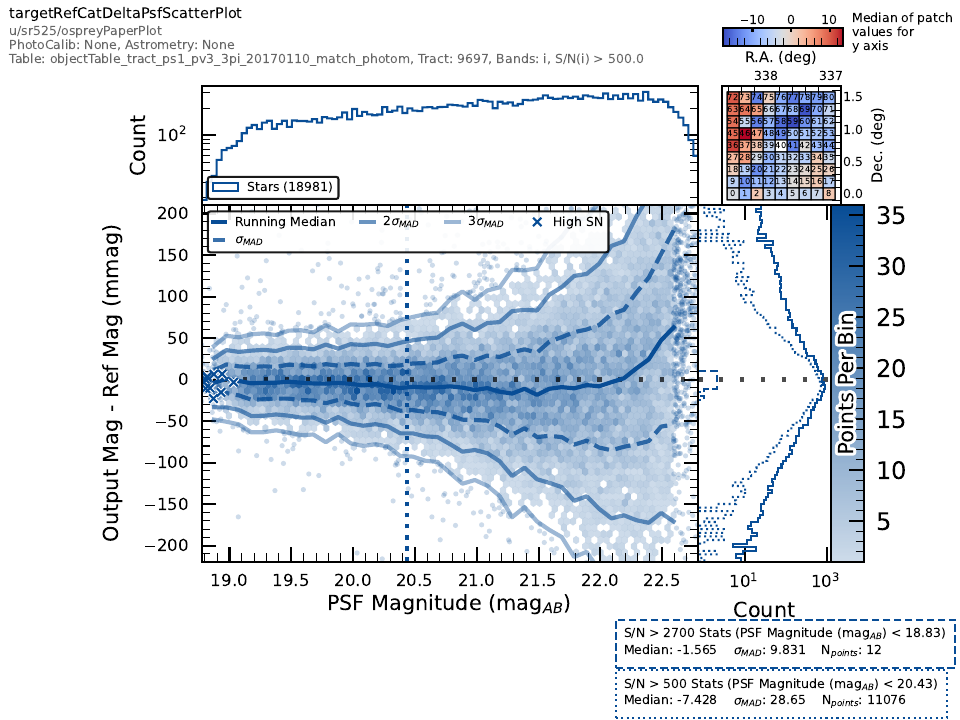}
\caption{An example figure produced as part of the standard processing by analysis tools. The plot is information-dense, as it is designed for an audience familiar with the outputs, but a simplified version can also be produced for talks and publications by setting a config option.
The metrics shown in the bottom right of the plot are also saved separately to be displayed by various pieces of QA tooling.}
\label{fig:atools}
\end{figure}

An example plot, made with HSC data, is shown in Fig.~\ref{fig:atools}.
The package was designed to handle the large data volumes and memory requirements that the survey will generate to ensure that the initial QA products required are rapidly made and readily available for fast action on any emergent data quality issues.
The individual tools run in the pipelines to calculate the metrics can then be reused in an interactive environment, such as a script or notebook, allowing further investigation of emerging issues using exactly what was originally run.

\texttt{analysis\_tools} is the successor to \texttt{faro} \citep{2022SPIE12189E..0MG}, \texttt{analysis\_drp}, and \texttt{MetricTask} from the \texttt{verify} package.
Previously the QA plots and metrics were calculated by different packages and run in different tasks; this allowed the code and selection criteria used to get out of sync between the plots and metrics, and in some cases it led to extreme duplication of I/O.
The more consolidated approach used by \texttt{analysis\_tools} improves memory and speed performance as well as maintaining consistency.

Further information and examples can be found in \citet{DMTN-314}.

% <<< input{sections/analysis/analysis_tools} <<<
% >>> input{sections/analysis/source_injection} >>>
\subsection{Source Injection}
\label{sec:source_injection}

\begin{figure}
    \centering
    \includegraphics[width=0.95\linewidth]{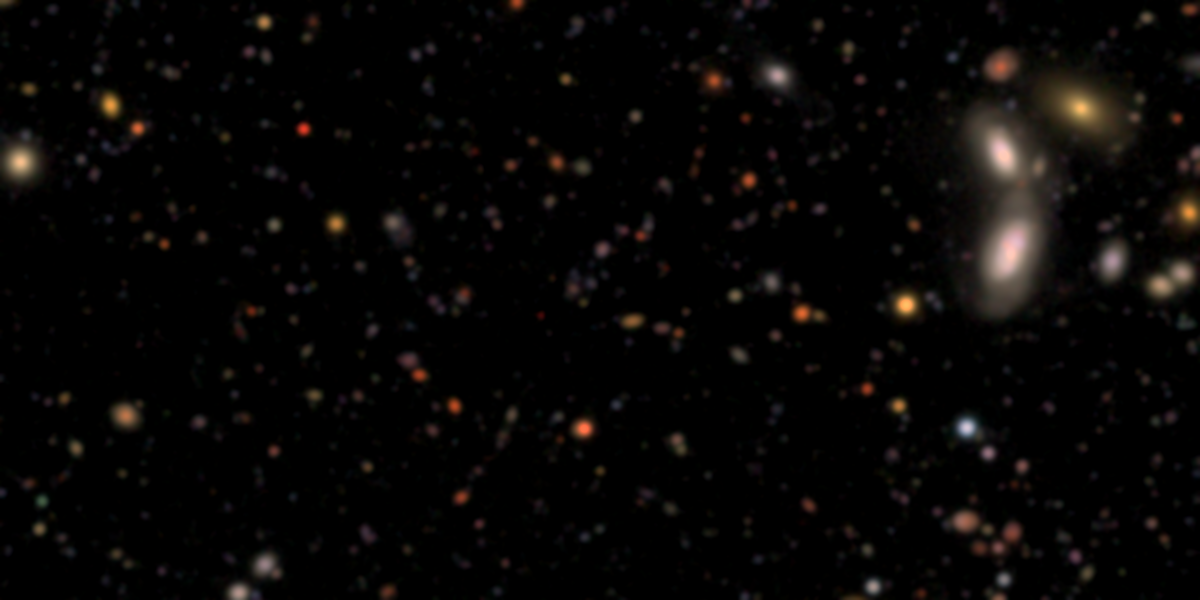}
    \includegraphics[width=0.95\linewidth]{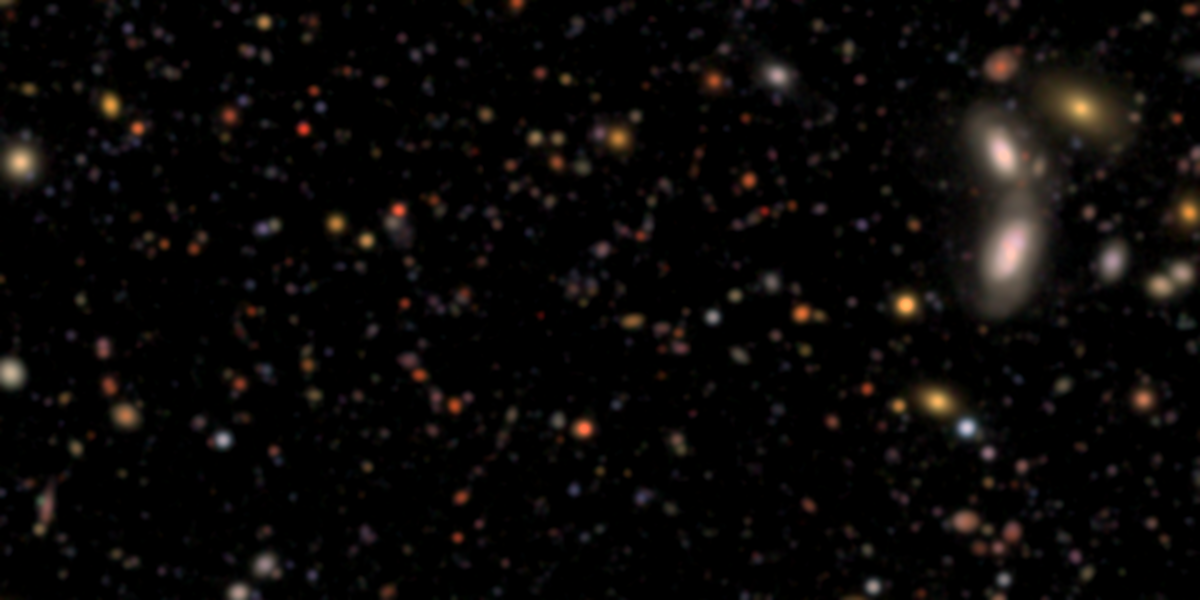}
    \includegraphics[width=0.95\linewidth]{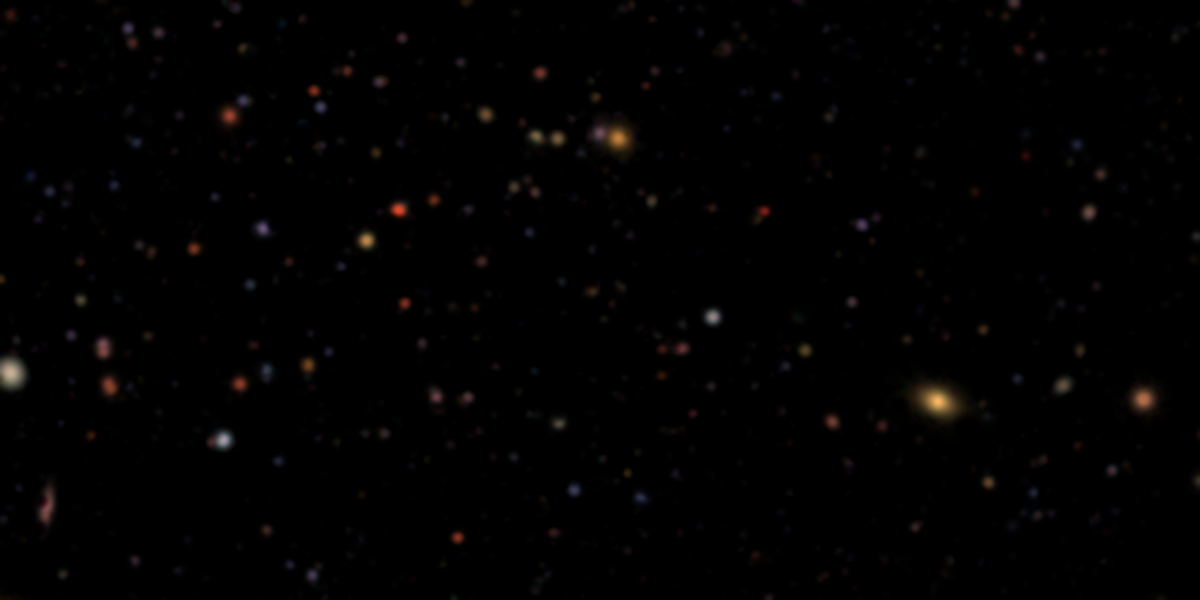}
    \caption{
      An LSSTComCam Data Preview 1 cutout from a deep coadd image \citep{10.71929/rubin/2570313} at tract 5063 patch 34, with sources injected from the DC2 catalog \citep{2021ApJS..253...31L}.
      The top panel is the original image and the middle panel shows the image with the injected sources.
      The bottom panel shows the difference between the two images.
    }
    \label{fig:source_injection_example}
\end{figure}

The \texttt{source\_injection} package contains tools designed to assist in the injection of synthetic sources into scientific imaging.
Source injection is a powerful tool for testing the algorithmic performance of the LSST Science Pipelines, generating measurements on synthetic sources where the truth is known and facilitating subsequent quality assurance checks.
Synthetic source generation and injection capability is provided by the \textsc{GalSim} software package \citep{2015A&C....10..121R}.
An example showcasing the injection of a series of synthetic sources from the DC2 catalog \citep{2021ApJS..253...31L} into an LSSTComCam coadd image from DP1 \citep{10.71929/rubin/2570313} is shown in Figure~\ref{fig:source_injection_example}.

Synthetic sources can be injected into any imaging data product output by the LSST Science Pipelines, including visit-level exposure-type or visit-type datasets (i.e., datasets with the dimension \texttt{exposure} or \texttt{visit}), or into a coadd-level coadded dataset.

Each task operates similarly: read in an injection catalog containing the parameters of the sources to be injected, generate sources using \textsc{GalSim}, and inject them into the input image.
An additional mask plane (\texttt{INJECTED} by default) is appended to the image mask to identify pixels that have been touched by injected sources.
Optional modifications to the noise profiles of injected sources and the variance plane of the image can also be performed.

Once source injection has completed, the source injection task will output two dataset types: an injected image, and an associated injected catalog.
The injected image is a copy of the original image with the injected sources added.
The injected catalog is a catalog of the injected sources, with the same schema as the original catalog and additional columns describing per-source source injection success outcomes.
% <<< input{sections/analysis/source_injection} <<<

% >>> input{sections/releases} >>>
\section{Pipeline Releases}

During the project construction phase we made releases approximately every six months.
We never formally used semantic versioning for these releases since there were many changes during the six months that would all inevitably correspond to a major release.
For people who wish to keep up with software developments we also make weekly releases available.
These releases are time-boxed and whatever state the software is in on the day of release is part of the release.
The main constraint on making a release is whether every CI package passes tests, and this is generally true because merges are gated on Jenkins builds being successful.
The community is always able to try the weekly releases with the caveat that they get much less testing and characterization than the formal releases.

During operations our plan is to make new formal software releases with each data release.
We will make the initial release branch during the first stage of production and backport fixes from our main branch as each stage is completed.
This means that when a data release is made the core of the associated pipeline software is many months older than the main development branch.

The large gaps between releases mean that normal software deprecation cycles where one warns in one release and then removes in a later release, leading to potentially two-year delays between deciding to deprecate and removing the code, are not really feasible.
Our main focus is supporting a data release and so while we do use deprecation cycles of a few months, we use the weekly releases to publicize deprecations in conjunction with announcements on the Rubin Community forum.

Algorithms are selected based on their speed, robustness, and performance against the Science Requirements Document (SRD) metrics \citep{LPM-17}.
These metrics are evaluated via regular regression tests.
A small test set (less than a square degree) is run through the alert and data release production pipelines nightly, a medium test set (about 10 square degrees) runs biweekly, and a data release production-sized test set (hundreds of square degrees) runs annually.

There are no controls on implementing new algorithms as pipeline tasks in the pipelines, as long as they meet our criteria in the development guide \citep{RubinDevGuide}.
However, there are strict controls over which algorithms are used by each of the alert production and data release production pipelines by default.
Before swapping one algorithm, configuration, or behavior change in as the DRP or AP pipeline default, the proposer must demonstrate that the change performs better on the SRD metrics, runtime metrics, and robustness (no runtime or algorithmic failures) on the biweekly-regression-test-sized dataset.
The proposer presents their proposal including the resulting metrics comparison at the Pipelines Team Meeting, on the pipelines Slack channel, or in an RFC.\footnote{See \url{https://developer.lsst.io/communications/rfc.html} for an overview of the RFC process.}
If the pull request is from the community and the author does not have access to this dataset or other Rubin resources, a DM team member will shepherd the PR through the controls and review process for them.
% <<< input{sections/releases} <<<
% This is a section.
% >>> input{sections/conclusions} >>>
\section{Conclusions}
\label{sec:conclusions}

The LSST Science Pipelines Software has been developed over 20 years to support the processing of the Legacy Survey of Space and Time.
It has been used to process formal data releases from both Hyper Suprime-Cam and the Rubin Observatory's LSSTComCam Data Preview 1 and LSSTCam Data Preview 2.
The software is designed to be extensible and reusable, supporting a plugin architecture that allows new algorithms to be added without modifying the core codebase and includes a dataset tracking system and graph builder that supports scaling of processing on large batch systems.
The success of this architecture is further demonstrated by its adoption by several other astronomical surveys, including VISTA's VIRCAM, WFST, GOTO, and SPHEREx, for processing their data.
The software will continue to be developed and evolve as new data releases are made and to support the community.
% <<< input{sections/conclusions} <<<

\begin{acknowledgments}
This material is based upon work supported in part by the National Science Foundation through Cooperative Agreements AST-1258333 and AST-2241526 and Cooperative Support Agreements AST-1202910 and AST-2211468 managed by the Association of Universities for Research in Astronomy (AURA), and the Department of Energy under Contract No.\ DE-AC02-76SF00515 with the SLAC National Accelerator Laboratory managed by Stanford University.
Additional Rubin Observatory funding comes from private donations, grants to universities, and in-kind support from LSST-DA Institutional Members.
This research has made use of NASA's Astrophysics Data System Bibliographic Services.
\end{acknowledgments}

\facilities{Rubin:Simonyi (LSSTCam), Rubin:1.2m (LATISS)}
\software{%
ndarray (\url{https://github.com/ndarray/ndarray}),
astropy \citep{2022ApJ...935..167A},
pytest \citep{pytest},
matplotlib \citep{2007CSE.....9...90H},
galsim \citep{2015A&C....10..121R},
numpy \citep{2020Natur.585..357H},
gbdes \citep{2022ascl.soft10011B,2017PASP..129g4503B},
Starlink's \citep{2022ASPC..532..559B} AST \citep{2016A&C....15...33B},
fgcm (\url{https://github.com/erykoff/fgcm}),
MODTRAN \citep{1999SPIE.3756..348B},
Sorcha \citep{2025AJ....170..100M},
mpsky \citep{mpsky},
CFITSIO \citep{2010ascl.soft10001P},
Piff \citep{2021ascl.soft02024J,2021MNRAS.501.1282J},
PSFEx \citep{2011ASPC..442..435B,2013ascl.soft01001B},
Boost (\url{https://www.boost.org/}),
Eigen \citep{eigenweb},
GSL \citep{galassi2009gnu},
}

\bibliography{PSTN-019_flattened}
\bibliographystyle{aasjournal}

\appendix
\twocolumngrid
% >>> input{sections/appendix} >>>
\section{Pipeline Configuration}
\label{sec:appendix_config}

An example of a pipeline configuration can be seen in the following code block which shows a fragment used to configure one of the shape measurement routines, edited slightly for clarity.
This abstraction of configuration from algorithm is critical for maintainability, allowing different processing runs to experiment with different configurations without having to edit any code.
It also provides a mechanism to deprecate configuration settings that will change in future versions of the software stack, allowing users an easy migration path.

\begin{minipage}{\columnwidth}
    \begin{lstlisting}[caption=Code configuration in Python, language=python]
pkg = "meas_extensions_shapeHsm"
uri = f"eups://{pkg}/config/enable.py"
try:
    config.load(uri)
except LookupError as e:
    print(f"Cannot enable shapeHSM {e}")
else:
    plugins = config.plugins
    plugin = plugins["ext_shapeHSM_HsmShapeRegauss"]
    plugin.deblendNChild = "deblend_nChild"
    # Enable debiased moments
    config.plugins.names |= ["ext_shapeHSM_HsmPsfMomentsDebiased"]
    \end{lstlisting}
\end{minipage}

The design of \texttt{pex\_config} centers around the concepts of \texttt{Field} and \texttt{Config} objects.
\texttt{Field}s represent individual configurable values---things like exposure times, image quality thresholds, or database connection strings.
Each \texttt{Field} is strongly typed, supporting a variety of data types such as integers, floats, strings, booleans, and lists.
\texttt{Config} objects, on the other hand, are containers that group related \texttt{Field}s together, creating logical units of configuration.
One of the highlights of \texttt{pex\_config} is its composability.
\texttt{Config} objects can be nested within other \texttt{Config} objects using a special \texttt{ConfigField}, allowing for the creation of complex, hierarchical configuration trees that mirror the structure of the pipelines themselves.
This allows for modularity and reuse of configuration components across different parts of the system.

A strength of \texttt{pex\_config} is its flexible application of configuration values.
Values can be set at multiple stages: via command-line arguments, loaded from configuration files, or defined directly within the pipeline code.
Importantly, these stages are applied progressively, with later stages overriding earlier ones.
This allows for a powerful combination of default settings, user-defined customizations, and dynamic adjustments.
Mechanisms also exist to apply values to all instances of a particular \texttt{Config} object within a tree, simplifying the management of shared parameters and ensuring consistency.

Beyond runtime configuration, \texttt{pex\_config} is deeply concerned with data provenance and reproducibility.
It provides mechanisms for persisting and restoring configuration values, allowing for complete tracking of pipeline parameters used in a particular data processing run.
Crucially, it also maintains a history of each \texttt{Field}'s value, recording when and where it was set---whether via the command line, a configuration file, or programmatically.
This detailed history is invaluable for debugging, auditing, and ensuring the reproducibility of scientific results.
The system also incorporates robust validation mechanisms, enabling checks on individual \texttt{Field}s and groups of values before they are used by the pipelines, preventing errors and ensuring data quality.
Validation can range from simple type checking, ensuring values fall within acceptable ranges or specific patterns, to complex custom functions that enforce specific constraints.

Finally, \texttt{pex\_config} is designed with documentation in mind.
All \texttt{Field}s and \texttt{Config} objects can be richly documented using documentation strings and attributes.
This documentation structure is not only readable by humans but can also be parsed by automated tools to generate comprehensive documentation pages, eliminating the need for manual documentation creation.
This ensures that the configuration system is well documented and easy to understand, even for new developers.

\section{Deblending}
\label{app:deblending}

\subsection{Single-band Deblending}
\label{sec:singleband_deblending}

Deblending on single-band images (i.e., visit images) is performed using the \texttt{meas\_deblender} package and is based on the deblender used in SDSS \citep{rhldeblend}, with a few differences that will be discussed shortly.
Similar to the SDSS deblender, the LSST deblender creates a template for each source in a blend using a very simple (yet computationally efficient) model for each peak position in a parent \texttt{Footprint}.
Once a template has been created for each peak in the blend, the deblender combines all of the source templates into a single blend model by summing their values in each pixel.
For each pixel in a source template, the ratio of the source template value to the total blend model is calculated and used to weight the pixel value from the image to create a model for each source.
The source models are thus flux-conserving in that adding them together will yield the original image except for pixels that do not appear in any of the individual templates.
A cleanup algorithm is then run to allocate the remaining pixels to one of the sources in the blend based on a set of criteria including distance to the center, brightness of the nearest sources, etc.

\subsection{Deblender Template Generation}

\begin{figure}
    \centering
    \plotone{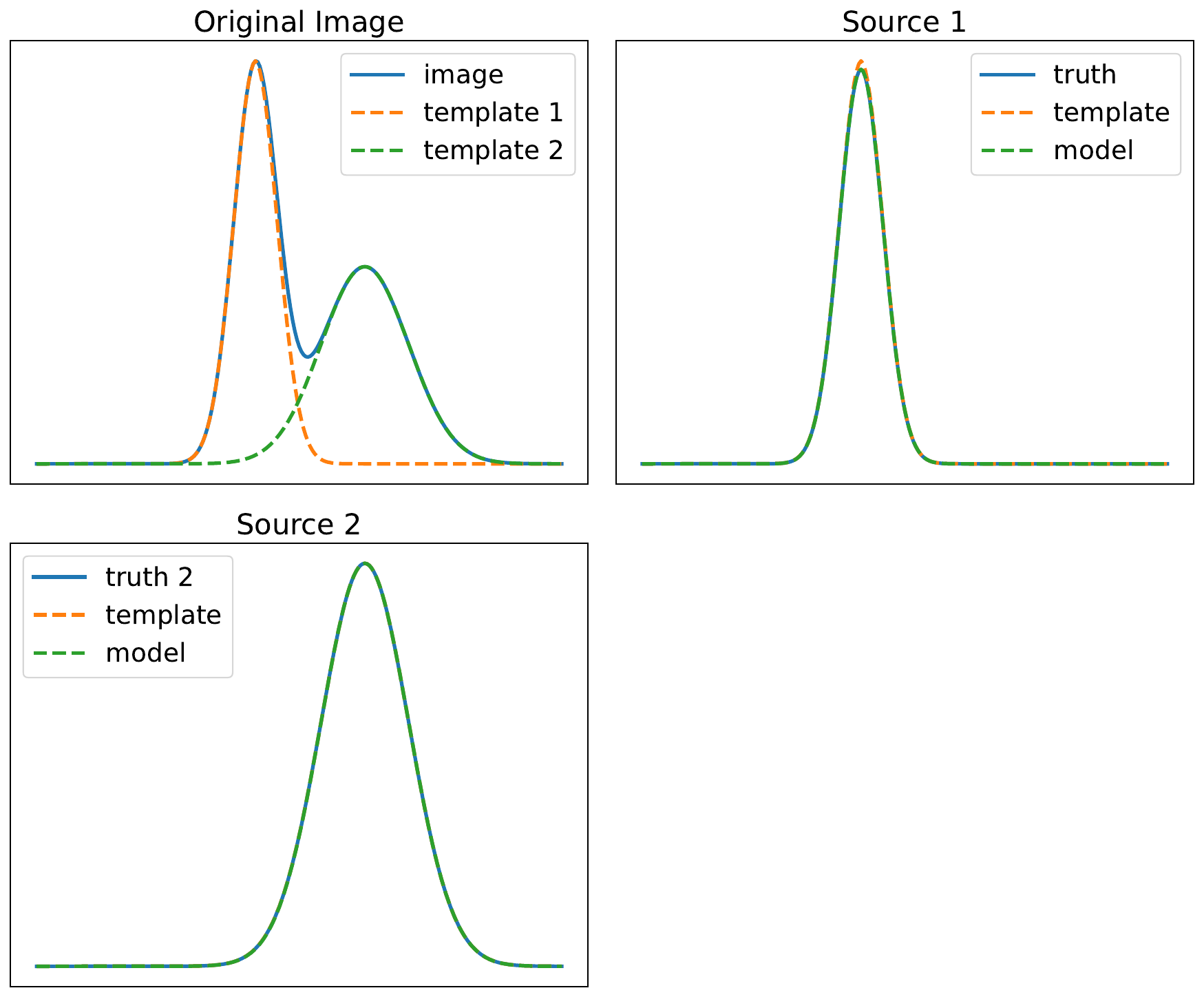}
    \caption{A 1D slice of two blended Gaussian sources illustrating how symmetry can be utilized to model blended sources.}
    \label{fig:simple_blend}
\end{figure}

The main ansatz of the SDSS deblending algorithm is that the flux from stars and galaxies in a ground-based telescope is nearly symmetric under a 180-degree rotation.
Figure \ref{fig:simple_blend} illustrates how a 1D slice through the center of two blended sources can exploit this symmetry by setting the pixels on opposite sides of the (integer) center pixel to the minimum value of both pixels.
In other words, for simple blends of only two sources the deblender can use the flux on the non-blended side to constrain the value of the flux on the blended side.
Despite the fact that stars (PSFs) and galaxies are not exactly symmetric, especially since their position is not exactly centered on a single pixel, this algorithm works quite well for generating templates in simple blends that very nearly approximate each source when redistributing flux from the image.

For sources with low signal-to-noise ratio, the algorithm fails due to noise in the image, generating galaxy templates that are typically very jagged and unphysical.
To combat this, for each \texttt{peak} in the parent \texttt{Footprint} the deblender first attempts to fit the flux from the image with a simple PSF model that allows its position, amplitude, and a linear background to vary.
If the fit has a reasonable $\chi^2$ value then the deblender will use this scaled PSF model as a template for the source.
Only for sources that cannot be adequately modeled with the PSF are the symmetric templates used.

\begin{figure}
    \centering
    \plotone{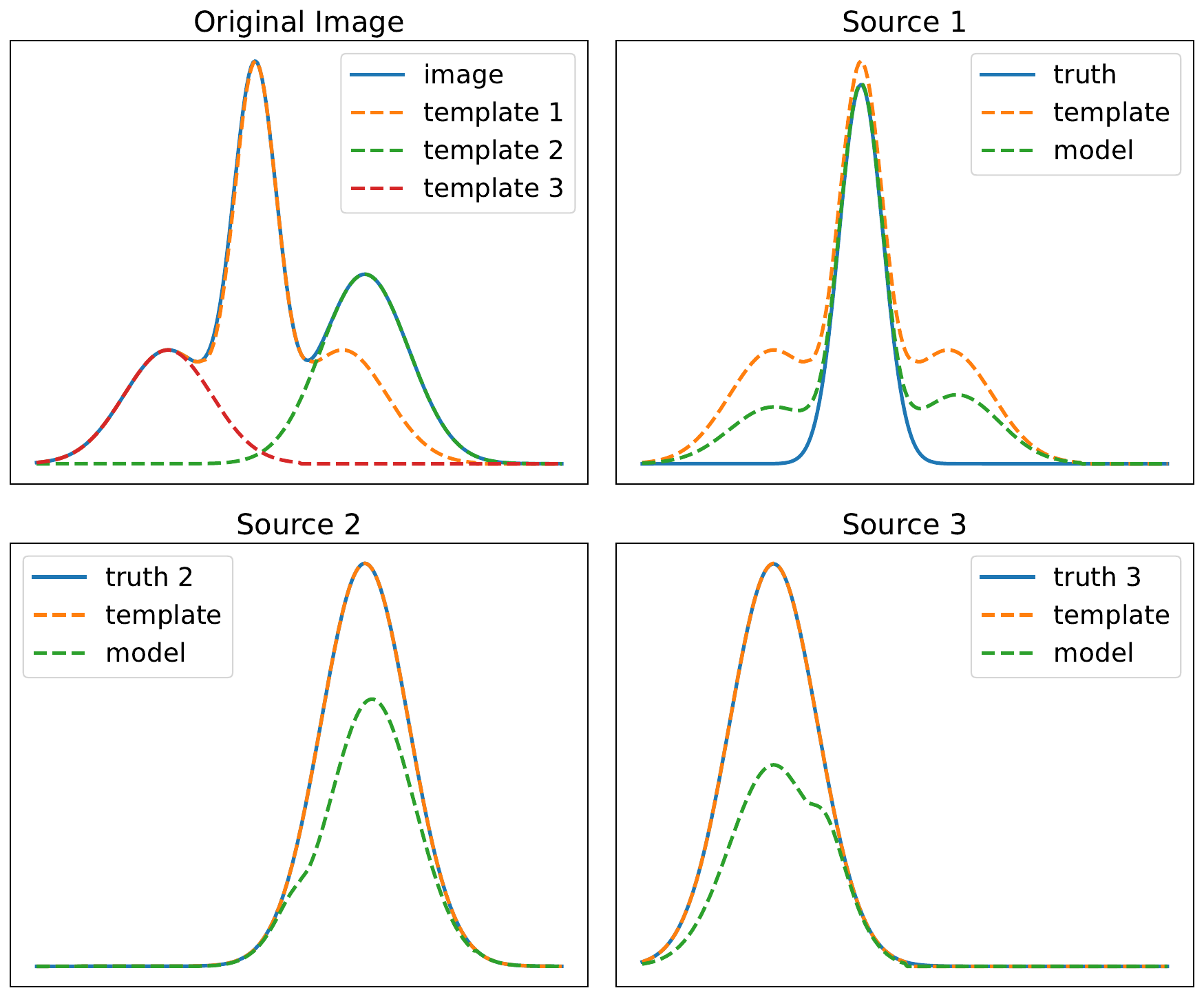}
    \caption{A 1D slice through three aligned Gaussian sources, demonstrating a failure case of relying on symmetry for generating deblender templates. Notice that for sources 2 and 3 the templates are reasonable but due to the inability of source 1 to use symmetry to constrain flux in the blended region, the resulting models for all three sources are poor. This catastrophic ``three in a row'' problem was part of the motivation for creating \textsc{Scarlet} to incorporate spectral information and a more rigorous iterative deblending algorithm.}
    \label{fig:complex_blend}
\end{figure}

The main failure point of this algorithm is when three (or more) sources lie along the same axis.
For example, Figure \ref{fig:complex_blend} illustrates a 1D slice through the center of three aligned sources.
In this case the minimum pixel on each side of the central source cannot constrain the flux at that radial location, which results in a template that has extra bumps from its neighbors.
This turns out to be more catastrophic than one might expect.
Notice that even the neighboring sources, which have very good templates created by using symmetry on their unblended side, have their resulting models contaminated due to the central object that steals flux from both of them.
In single visits the number of ``three in a row'' blends is small enough that we sacrifice the quality of the models for efficiency and still use the single-band deblender.
For LSST-depth coadds this becomes a significant problem, as deep coadds can have as much as 40\% of blends having three or more sources and a more sophisticated algorithm is needed.

\begin{figure*}[htb]
  \centering
  \includegraphics[width=\linewidth]{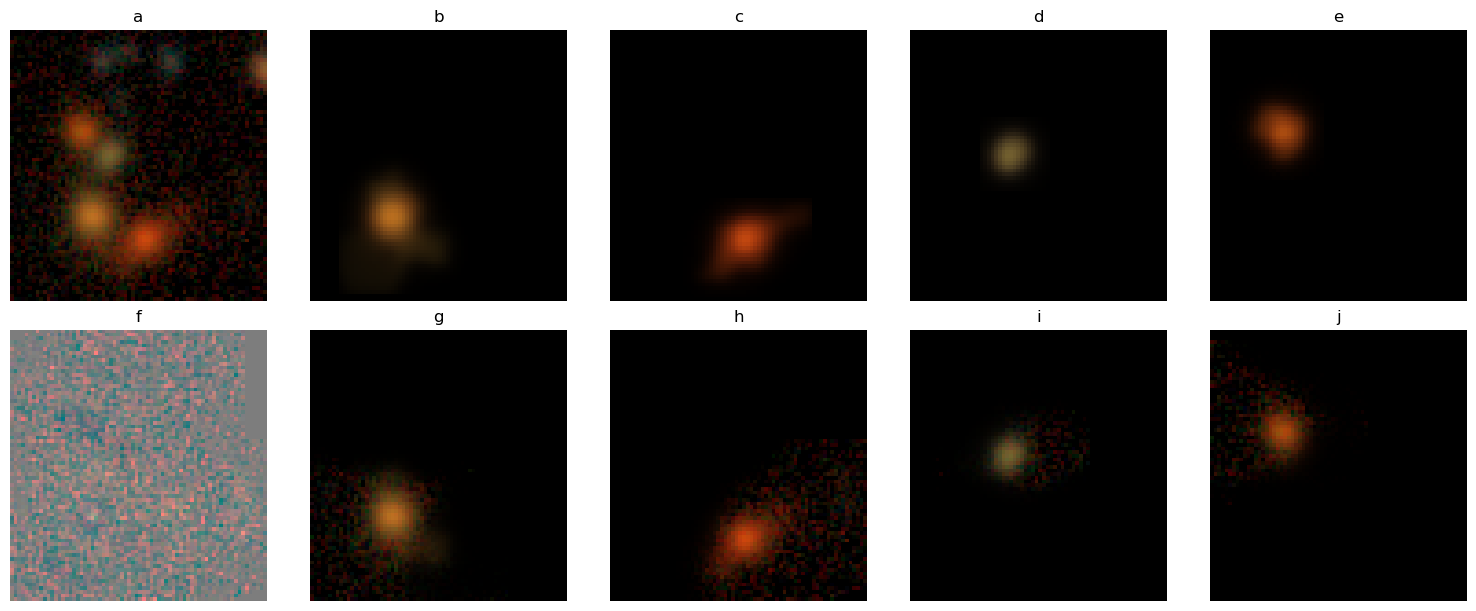}
  \caption{Example multi-object blend. The top row shows the observed multiband image (panel a) followed by the scarlet lite models of the four brightest sources (panels b through e). The bottom row shows the residual image when the scarlet lite models are subtracted from the observation (panel f) followed by the flux-conserved models (panels g through j), created by using the scarlet lite models as templates to redistribute flux from the observation so that the total flux is conserved. The flux-conserved models, injected in a noise-replaced image, are what the Science Pipelines use for measurement.}
  \label{fig:sample_blend}
\end{figure*}

\subsection{Multi-band Deblending}
\label{sec:multiband_deblending}

The multi-band deblender is an implementation of the \textsc{Scarlet} deblending algorithm described in \citet{2018A&C....24..129M}.
In our implementation, \texttt{scarlet\_lite}, we have made our own set of simplifying assumptions that are different from the original \textsc{Scarlet} algorithm to make it more efficient when used in a large ground-based survey like LSST.
Similar to the original \textsc{Scarlet}, we make the assumption that astrophysical objects can be thought of as a collection of components, where each component has the following properties:
\begin{itemize}
    \item Components have a single color (spectrum) that is the same in all pixels over its shape (morphology)
    \item Components have flux that monotonically decreases from the center
    \item Component flux is additive
\end{itemize}

The classic example is decomposing a single galaxy into bulge and disk components, where both the bulge and disk share a common center but have different spectra and morphologies.
Something more complex, like a grand-design spiral, could in theory be modeled as a source with multiple components, where spiral arms and star-forming regions could still be thought of as separate monotonic components.
For the Science Pipelines, we ignore those more complicated structures, as detection typically already shreds large galaxies into multiple sources.
Instead, we use a signal-to-noise cut where low-flux sources are modeled with a single component and higher-flux sources are modeled with two components.

Scarlet lite initializes models with nearly the same templates as those generated by the single-band deblender.
Using a $\chi^2$-like monochromatic image created by weighting each band by its inverse variance, scarlet lite creates initial morphology models that are symmetric from the center in the monochromatic image, with the additional constraint that the flux is monotonically decreasing from the center.
In order to satisfy the constraint that all pixels in the morphology have the same spectrum, scarlet models exist in a partially deconvolved frame with the seeing of a well-sampled but narrow Gaussian.
The initial spectrum of each source is determined using a least-squares fit of each monochromatic morphology, convolved with the difference kernel in each band to match the image, for each component.
It then uses proximal-\textsc{Adam} \citep[\textsc{PAdam};][]{2019arXiv191010094M} to iteratively update the spectrum and morphology with the given constraints until convergence or a maximum number of iterations is reached.
It should be noted that although we do use symmetry to initialize the scarlet models, we do not implement a symmetry constraint and the final models are not guaranteed to be symmetric.
The models are stored as the \texttt{object\_model\_data} data product, which contains all of the blends for a single patch.
Like the single-band deblender, the \texttt{scarlet\_lite} models are only used as templates to redistribute flux from the image and all measurements are performed on the flux-redistributed models.
An example of the deblender output for a DP2 field is shown in Figure \ref{fig:sample_blend}.
% <<< input{sections/appendix} <<<

\end{document}